\documentclass[prd,nofootinbib,preprintnumbers]{revtex4}
\usepackage{latexsym,amsmath,amssymb,amsbsy,graphicx}
\usepackage[cp1251]{inputenc}
\usepackage[T2A]{fontenc}
\usepackage{braket}
\usepackage{ulem}
\usepackage{xspace}
\usepackage{color}
\numberwithin{equation}{section}

\renewcommand{\sc}{\mathrm{sc}}
\renewcommand{\em}{\mathrm{el}}

\newcommand{\bs}{\beta'}

\newcommand{\cd}{\,\! ;  \,}

\newcommand{\ds}{\displaystyle}
\newcommand{\comment}[1]{}

\newcommand{\hsp}{\hspace{0.15em}}
\newcommand{\nhsp}{\hspace{-0.12em}}
\newcommand{\smhsp}{\hspace{0.06em}}

\newcommand{\kn}{{\kern 1pt}}
\newcommand{\akn}{{\kern -1pt}}

\newcommand{\ir}{r}

\newcommand{\e}{\mathrm{e}}

\newcommand{\tq}{\tilde{q}}
\newcommand{\te}{\tilde{\eta}}
\newcommand{\tx}{\tilde{x}}

\newcommand{\dDp}{d^{\hsp D}\nhsp p}
\newcommand{\dDq}{d^{\hsp D}\nhsp q}

\newcommand{\ddp}{d^{\smhsp d}\nhsp  p}
\newcommand{\ddq}{d^{\smhsp d} \nhsp  q}
\newcommand{\ddx}{d^{\kn d} \akn  x}

\newcommand{\pp}{\, ... \,}

\newcommand{\vp}{\vphantom{\frac{a}{a}}}

\newcommand{\nn}{\nonumber}

\newcommand{\reg}{{\rm reg}}
\newcommand{\ren}{{\rm ren}}

\newcommand{\mi}{\, | \, }

\newcommand{\ee}{\end{equation}}
\newcommand{\be}{\begin{equation}}

 \numberwithin{equation}{section}

\begin{document}

\title{Vacuum polarization and classical self-action near higher-dimensional defects}
\author{Yuri V. Grats and
Pavel Spirin\footnote{E-mails: grats@phys.msu.ru,
pspirin@physics.uoc.gr. }} \affiliation{Department of Theoretical
Physics, Faculty of Physics, Moscow State University, Moscow,
119991,  Russian Federation}



\begin{abstract}
We analyze the gravity-induced effects associated with a massless
scalar field in a higher-dimensional spacetime being the tensor
product of $(d-n)$-dimensional Minkowski space and $n$-dimensional
spherically/cylindrically-symmetric space with  a solid/planar
angle deficit. These spacetimes are considered as  simple models
for a multidimensional global monopole (if \mbox{$n\geqslant 3$})
or cosmic string (if \mbox{$n=2$}) with \mbox{$(d-n-1)$} flat extra
dimensions.    Thus, we refer to them as conical backgrounds. In
terms of the angular deficit value, we derive the perturbative
expression for the scalar Green's function, valid for any
\mbox{$d\geqslant 3$} and \mbox{$2\leqslant n\leqslant  d-1$}, and
compute it to the leading order. With the use of this Green's
function we compute the renormalized vacuum expectation value of
the field square $\langle \phi^{\kn 2}(x)\rangle_{\ren}$  and the
renormalized vacuum averaged of the scalar-field's energy-momentum
tensor $\langle T_{M N}(x)\rangle_{\ren}$ for arbitrary  $d$ and
$n$ from the interval mentioned above and arbitrary coupling
constant to the curvature $\xi$.

In particular, we revisit  the computation of the
vacuum polarization effects for a non-minimally coupled massless
scalar field in the spacetime of a straight cosmic string.

The same Green's function enables to consider the old purely
classical problem of the gravity-induced self-action of a
classical pointlike scalar or electric charge, placed at rest at some
fixed point of the space under consideration.

To deal with divergences, which appear in consideration of the
both problems, we apply the dimensional-regularization technique,
widely used in quantum field theory (QFT). The explicit dependence
of the results upon the dimensionalities of both the bulk and
conical submanifold, is discussed.
\end{abstract}

\maketitle


 \section{Introduction}

Through the last decades the higher-dimensional generalizations of
known four-dimensional solutions in General Relativity (GR)
became
 the object of intense research
in the context of widely developing higher-dimensional  theories.
It is enough to mention the possibility of the mini-black-hole
creation in the high energy physics experiments  \cite{BH}.
Experimental confirmation of such a creation is considered as one
of  tests on the existence of extra dimensions, or it has to set
new bounds on the parameters of the multidimensional theories
predicting the existence of mini-black-holes. Though at present,
there are no confirmations of the extra-dimension existence
\cite{CMS}, the modern theories stimulated the research of  the GR
in
 \mbox{$d>4$} spacetime dimensions. This implies not only the search of new solutions, but also the research of the
 higher-dimensional generalizations of the known four-dimensional solutions. The partial goal of such research
 is  to clarify, which predictions by GR are proper for four dimensions only, and which ones are universal and
 extended to higher dimensions. At the other hand, it is expected that the research of higher-dimensional
 generalizations allows to shed light on some peculiarities  of
 the standard four-dimensional theory and assists in the better understanding of the latter.
This research assumes not only the study of geometric features of higher-dimensional solutions, but also the
study of particularities
of the classical/quantum matter dynamics on their background.

The standard problems of research within the field theory on the
curved background, to which the physicists return through decades, are
the effects of the induced by gravity vacuum polarization and the
problem of self-action of the classical charged particle. These
two problems, weakly related at the first glance, in fact have a
number of common features. The main of those is that the both
problems are determined by the appropriate  Green's function being the solution  of partial differential
equation, which is sensitive to the global structure of the
manifold. Thus, the both effects become essentially non-local.
 Furthermore, for
the elimination of divergences arising in the both cases, one uses
the same techniques.

 The present work is devoted to the consideration of gravity-induced effects of the
 vacuum polarization of a massless scalar field and the self-action of a scalar or electric
 charge on the ultrastatic spacetime being the   product of \mbox{$(d-n)$}-dimensional Minkowski spacetime and
 $n$-dimensional spherically-symmetric space with an angular deficit.

 We will be concentrated on the computation
 of the renormalized vacuum expectation values (VEV)
for $\langle \phi^{\kn 2}(x)\rangle_{\ren}$ and $\langle \kn
T_{MN}(x)\rangle_{\ren}$, as well as of calculation of the
renormalized self-energy $U_{\ren}(x)$ and self-force
$F_{\ren}(x)$ of the static scalar or electric charge. For the
regularization of formally diverging expressions we will use the
dimensional-regularization technique.



The paper is organized as follows: Introduction is the first
section. In the Second section, the Setup, we briefly present the
background metric with angular deficit in arbitrary spacetime
dimension and derive the initial expressions  for the subsequent
computation of classical self-force and  vacuum averages. The
perturbation theory we use, is described in the Section
\ref{PertTheo}, where we also construct the approximated Green's function.
The Section \ref{phi2ren} is devoted to the computation
 of  renormalized vacuum averaged $\langle\phi^{\kn 2}(x)\rangle_{\ren}$
 in the dimensional-regularization scheme. The comparison with
 the analogous results known
  in the literature, is presented.
The renormalized stress-energy tensor is computed in the Section
\ref{EMTVEV}. The classical self-energy and self-force of a
pointlike scalar or electric charge in the spacetime-at-hand, are
computed in the Section~\ref{Self-force}. In the Section
\ref{Cosmostring} we discuss the special case of an infinitely
thin cosmic string. We show that there is a some ambiguity it the previous
calculations and propose an alternative approach to the problem.
In the last Section \ref{Concl}, the Conclusion, we summarize the
results and prospects. Useful integrals are given in the single
Appendix.

We use the  units  $G=c=\hbar =1$ and metric with the signature
$(-, +, +, \pp, + )$.

\section{Setup}\label{Setup}
 In the model we consider quantized or classical massless  scalar field $\phi$, living in
the static \mbox{$d-$}dimensional bulk with $n$\kn{}dimensional
submanifold with solid or planar angular deficit. This
$n-$dimensional subspace may be considered as created by the $n-$dimensional
global monopole
 (for \mbox{$n\geqslant 3$}) or as a
straight cosmic string (for \mbox{$n=2$}).

First we overview the background geometry.

\subsection{Background of the cosmic string and the global monopole, and their
higher-dimensional analogues}

The metric of a straight infinitely thin cosmic string with a mass
per unit length $\mu$, located along the $z-$axis in four
spacetime dimensions, in cylindric coordinates reads
\be \label{Metstr}
ds^2=-dt^2+dz^2+d\rho ^2+\beta ^2{\rho }^2{d\varphi }^2\ ,
\ee
where $\beta=1-4G\mu$\kn. (For the review of the formation,
evolution and geometry of topological defects and some physical
effects near them see \cite{Form,Top} and Refs therein). The
corresponding Riemann tensor vanishes everywhere except
 the symmetry axis $\rho=0$, where it has a $\delta $-like
singularity~\cite{Sokolov77}. Straight string does not affect the
local geometry of the spacetime, its effect on matter fields is
purely topological, and the dimensionless parameter $G\mu $ is the
only parameter which measures the effect of conical structure on
the dynamics of classical and quantized matter.

In some applications it is more appropriate to use coordinates
$(t,x,y,z)$, which are conformally Cartesian on the plane
transverse to the string. With the radial-coordinate
transformation $\rho \to r$ as
$$
 \rho =\frac{r_{0}}{\beta}\Bigl( \frac{r}{r_{0}}\Bigr) ^{\!\beta }\ ,\qquad
x^1=r\cos \varphi ,\quad x^2=r\sin \varphi \ ,
$$
where $r_{0}$ is an arbitrary scale with the length
dimensionality, the line element~(\ref{Metstr}) takes the form
\begin{eqnarray}\label{Metstr3}
ds^2=-dt^2+dz^2+\e^{-2(1-\beta )\ln (r/r_{0})}\delta
_{ab}\,dx^{a}dx^{b}\,,
\end{eqnarray}%
where $r^2=\delta _{ab}\,x^{a}x^{b}$\ ,$\quad a,b=1,2\,.$

The idea to use the conformal coordinates was put forward in the
framework of a low-dimensional gravity~\cite{DJH}. In this case it
gives the possibility to find a self-consistent solution for the
metric of a multi-center space, i.e. a static $(2+1)$--dimensional
spacetime of $N$ point masses. Later it was shown that the line
element of a multi-center spacetime
can be generalized for the case of $N$ parallel cosmic strings~\cite{Letelier87}. The same idea enables to obtain the explicit
solutions of the problem of topological self-action in the
multicenter and multistring
spacetimes~\cite{Grats96,Grats98_1,Grats98,Grats00}, and provides
an appropriate framework for consideration of the vacuum
polarization effect in the spacetime of multiple cosmic strings
and in particular, the vacuum Casimir-like interaction of parallel
strings~\cite{Grats1995}.

One can consider the generalization of the metric (\ref{Metstr}) and (\ref%
{Metstr3}) for a spherically symmetric case, when any plane
containing the center of symmetry and dividing the space into two
equal parts is a cone with the angular deficit $\delta\varphi=2\pi (1-\beta )$
\begin{equation}\label{Metmon2}
ds^2=-dt^2+d\varrho^2+{\beta }^2\varrho^2(d\theta^2+\sin^2\!\theta\,
d\varphi ^2)\,.
\end{equation}%
This metric describes an ultrastatic spherically symmetric
spacetime with the solid angle deficit equal to~ $4\pi (1-\beta
^2)$.

Expression~(\ref{Metmon2}) approximates the metric of a global
monopole~\cite{Vilenkin89,Harari90}. Strictly speaking, the metric
of a global monopole contains a mass term, but this term is too
small to be of importance on astrophysical scale.

As in the string case, there is a possibility to use
conformally Cartesian coordinates on the section $t={\rm {const}}$
of the spacetime (\ref{Metmon2}). After redefinition of the radial
coordinate $\beta \varrho=r_{0}(r/r_{0})^{\beta }$ the metric of
the spatial sector of the above line element takes the
conformally Euclidean form. Thus, we can introduce a set of
Cartesian coordinates $\{x^{i}\}\ ,i=1,2,3$\thinspace\ with
usual relation with the spherical coordinates
$r,\theta ,\varphi $. In these coordinates metric~(\ref{Metmon2})
reduces to the form
\begin{eqnarray}\label{Metmon1}
ds^2=-dt^2+\e^{-2(1-\beta )\ln (r/r_{0})}\delta
_{ik}dx^{i}dx^{k}\,,
\end{eqnarray}
where  $r^2=\delta _{ik}\kn x^{i}x^{k}$\,,$\quad i,k=1,2,3$.

We see that both conical defects have no Newtonian potential and
exert no gravitational force on the surrounding matter. For both
defects their gravitational properties are determined by the
deficit angle only. The main difference of a global monopole from
the case of a cosmic string is that the monopole spacetime is not
locally flat, and its gravitational field provides a tidal
acceleration which is proportional to~$r^{-2\beta}$.

Below we will consider multidimensional generalization
of the spaces (\ref{Metstr3}) and (\ref{Metmon1}), with arbitrary number of conical and flat spatial dimensions. The corresponding
metric reads:
\begin{align}\label{Metnew}
ds^2 \equiv   g_{MN}\,dx^M dx^N=-dt^2+dx_{d-1}^2+\pp +dx_{n+1}^2 +
\e^{-2(1-\beta)\ln r}\,\delta _{ik}dx^{i}dx^{k}\,,
\end{align}
with $r^2  \equiv \delta _{ik}x^{i}x^{k}$ and \mbox{$i,k, \pp =
{1, \pp, n}$} while \mbox{$M,N, \pp = 0, 1, \pp, d-1$}. Here
\mbox{$d\geqslant 3$} and \mbox{$2\leqslant n \leqslant d-1$}.
Without loss of generality we put $r_0$ equal to unity.


The spacetime with metric (\ref{Metnew}) represents the tensor
product of the $(d-n)$-dimensional Minkowski space and the
$n$-dimensional  centro-symmetric conformally flat space with a
solid angle deficit equal to
\mbox{$\delta\Omega=2\kn(1-\beta^2)\kn\pi^{n/2}/\Gamma\left(n/2\right)$},
if $n\geqslant 3$, or planar angular deficit equal to
\mbox{$\delta \varphi=2\pi (1-\beta)$}, if \mbox{$n=2$}.

The corresponding Ricci tensor and the scalar curvature are determined
by the conical sector only:
\begin{align}\label{Riccies}
&R_{ik}=2\pi(1-\beta)\,\delta^2(\mathbf{r})\,\delta_{ik}\, ,
& & R=4\pi(1-\beta)\,r^{2(1-\beta)}\,\delta^2(\mathbf{r})\, ,& & n=2; \nn  \\
& R_{ik}=(1-\beta^2)(n-2)\frac{r^2\,\delta_{ik}-x_i\,x_k}{r^4}\,
, &  & R=(1-\beta^2)\frac{(n-1)(n-2)}{r^{2\beta}}\, ,& &
n\geqslant 3\,.
\end{align}

For these spaces and corresponding Green's functions we will use
the notations $(d, n)$ and \mbox{$G(x, x'\mi d, n)$}. Notice, in
these notations, the spacetime  of a straight infinitely thin
cosmic string and that one of a point global monopole in four
spacetime dimensions have the type $(4, 2)$ and $(4, 3)$,
respectively.

So, (\ref{Metnew}) represents the multidimensional generalization of
the four-dimensional solutions obtained in
\cite{Vilenkin_81,Gott_85} and \cite{Vilenkin89,Harari90},
correspondingly.

For the first time metric of the form (\ref{Metnew}) with
two-dimensional conical subspace ($n=2$) was considered in the
paper \cite{Dowker_1987}. Later a number of solutions for a
coupled system of the Einstein equation and the equations of
motion for $n$ scalars was found and analyzed in~\cite{Vilenkin_2000}. It was shown, that the \mbox{$n\geqslant 3$} solution
with equal-to-zero cosmological constant  has approximately the
form (\ref{Metnew}) (in our coordinates).
Thus, the metric
(\ref{Metnew}) describes the conical defects which live in a
$d-$dimensional bulk, having a flat \mbox{$(d-n-1)$}\kn-brane as a core.
Some tiny QFT effects have been found on these backgrounds for
some particular dimensionalities of the bulk dimension $d$ and the
dimension of the conical subspace $n$. The vacuum polarization effects
for a massless scalar and fermionic fields on the
higher-dimensional monopole/string spacetime were investigated in
\cite{Bezerra_2002,Bezerra_2006} and
\cite{Dowker_1987,Grats1995,Spinelly_2008,Grats2016}. In \cite{Bezerra_2010}
the authors analyze the vacuum fluctuations of a quantum bosonic
and fermionic currents induced by a magnetic flux running along
the string. In this paper we continue the investigation of quantum
and classical field-theoretical processes on the generalized
background (\ref{Metnew}).

The geometry of the spacetime under consideration is simple enough
and the metric does not contain any dimensional parameters.
Nevertheless we cannot compute explicitly Green's function
\mbox{$G(x, x'\mi d, n)$} in a workable closed form. So, we
restrict our consideration by the particular case of a small
angular deficit; in what follows, we put  $(1-\beta)\ll 1$\,.
It enables us to obtain perturbatively the universal expression for
the Green's function, which is valid for any $d$  and $n$ and for any
value of the coupling constant $\xi$.

\subsection{Self-energy of a pointlike charge in a static spacetime: formalism}

Let us consider a massless scalar field $\phi$ with a source
$j$ in a static $d-$ dimensional spacetime with the metric
\begin{equation} \label{e1} ds^2\equiv   g_{MN}\,dx^M dx^N=g_{00}\,dt^2+g_{\mu
\nu}~dx^{\mu} dx^{\nu}\, ,\qquad g_{00}<0\, .
\end{equation}
In this subsection the small Greek indices $\mu,\nu,\pp$ run over
all spatial coordinates $1,2,\pp,d-1$.

The interaction of   scalar field with the bulk curvature $R$ is
introduced via coupling $\xi$, while interaction with charges is
introduced by the charge density $j(x)$ in a standard way:
\begin{align}\label{action1}
S_{\rm tot}=-\frac{1}{2}\int \ddx  \sqrt{-g} \left(\phi_{\kn \cd
M}\,\phi^{\kn \cd M} +\xi R \phi^2- 2 \phi \kn j \right)+S_j\, .
\end{align}
$S_j$ is the action for a charged matter.

From (\ref{action1}) one derives the equation of motion for scalar
field:
\begin{equation}\label{e2}
  \partial_{M}\left(\sqrt{-g}\, g^{MN}
\partial_{N}  \phi\right)-\xi \sqrt{-g} \, \phi  R = -
\sqrt{-g}\,j\,
\end{equation}

In the static case, when $\partial_0\phi=0=\partial_0 g_{MN}$, and
pointlike charge $q$ placed at a fixed  spatial point $x$ it reads
\begin{align}\label{e2_sc2}
  \partial_{\mu}\left(\sqrt{-g}\, g^{\kn \mu\nu}
\partial_{\nu} \phi \right)-\xi R \kn  \phi = - \sqrt{-g}\,j\,
\end{align}
where
\begin{equation}\label{source}
j(x')=q \frac{\delta^{d-1} \kn( x- x')}{\sqrt{-g}}\,.
\end{equation}

The field energy  in a static spacetime reads
\begin{equation}\label{Dzuba8}
  U=  - \int
T_{0}^{0}\,\sqrt{-g}\,d^{\kn d-1} x\, ,
\end{equation}
where $T_{0}^{0}$ stands for zero-zero component of the
energy-momentum tensor, which  for the scalar field  is derived from the action
(\ref{action1}) and given by
\begin{align}\label{EMT1}
T_{M}^{N}=  (1-2\xi) \, \phi_{, M}\phi^{\,, N}  + \frac{
4\xi-1}{2}\, \phi_{, L} \phi^{\, , L}  \,\delta_{M}^{N}-2\kn\xi
\phi_{\kn ; M}{}^{; N} \phi  + 2\xi \phi\,  \Box\phi
\,\delta_{M}^{N}+\frac{ \xi}{2}\,\bigl(2 R_{M}^{N} -R
\,\delta_{M}^{N}\bigr)\, \phi^{\kn 2} \,.
\end{align}

Note, that the interaction part of the action does not
contribute to the field energy-momentum tensor. It is particularly
obvious in the case under consideration since for a pointlike charge
with the source (\ref{source})  the Lagrangian density reads  $
{\cal{L}}_{\rm int}=\sqrt{-g}\,\phi\, j=q\,\delta^{d-1}(x-x')$ and
does not depend on the metric.

Making use the fact that the field and the metric are static  we
have
$$T_{0}^{0}=   - \frac{1- 4\kn \xi}{2}\,  \phi_{,\kn  \mu} \phi^{\,,\kn \mu}  + 2\kn \xi \phi \frac{1}{\sqrt{-g}}\,
\partial_{\mu}\Bigl[\sqrt{-g}g^{\mu\nu}\partial_{\nu}\phi\Bigr] +\xi\biggl(R_{\,0}^{\,0}-\frac{1}{2}R\biggr) \,
\phi^{\kn 2} \,.$$
Substituting $T^{0}_{0}$, the  scalar-field energy is given by

\begin{equation*}
U_{\sc}= \frac{1-4\xi}{2 }\int d^{\kn d-1} x\, \partial_{\mu}\left(
\sqrt{-{g}}\, g^{\kn \mu\nu}   \phi\,\partial_{\nu}  \phi  \right)
- \frac{1}{2 }\int d^{\kn d-1} x\, \Bigl[ \phi\,
\partial_{\mu}\left(\sqrt{-{g}}\, g^{\kn \mu\nu}\,
\partial_{\nu}  \phi\right) +\sqrt{-g} \xi \left(2R_{\,0}^{\,0}-R\right)\, \phi^{\kn 2}\Bigr]\,.
\end{equation*}
and  integrating with  help of the Gauss' theorem,  only the
second integral survives. Simplifying and taking  help of the
field equation (\ref{e2_sc2}), (\ref{Dzuba8}) becomes
\begin{equation}\label{qqq0_sc}
U_{\sc}= \frac{1}{2}\int d^{\kn d-1} x\, \sqrt{-{g}}\, \left[ \phi\kn
 j-2\xi\,R_{\,0}^{\,0}\,\phi^{2}\right]\,.
\end{equation}

The corresponding form   via   Green's function of the
Eq.\,(\ref{e2_sc2}) reads:
\begin{equation*}
U_{\sc} =  \frac{1}{2}\int d^{\kn d-1} x\,d^{\kn d-1}
x'\,\sqrt{g(x)\,{g}(x')}\,j(x)\,G(x, x')\,j(x')-\xi\int d^{\kn
d-1} x\, \sqrt{-{g}}\,R_{\,0}^{\,0}\,\phi^{2}\,\,,
\end{equation*}
where $G(x, x')$   satisfies
\begin{equation}\label{e5}
 \partial_{\mu}\left(\sqrt{-{g}}\, g^{\kn \mu\nu}\hsp\partial_ {\nu} G(x, x')\right)
  -\xi R \sqrt{-{g}} \,G(x, x')
  =- \sqrt{-{g}}\,\delta^{\kn d-1}(x, x')\,.
\end{equation}

Thus for a point charge localized at the point $x$ of the
spacetime from Eg.(\ref{source}) we get
\begin{equation}\label{e6}
 U_{\sc}(x)= \frac{1}{2}\, q^2\,G(x,x) -\xi\int d^{\kn d-1} x\,
\sqrt{-{g}}\,R_{\,0}^{\,0}\,\phi^{2}\,\, .
\end{equation}

Note, that  for a general case of a static spacetime
one has $g=g_{00}\det(g_{\mu\nu})$, while $R$ in Eq.(\ref{e5})
stands for the scalar curvature of the whole $d$\kn-dimensional
space.

In addition,  if the spacetime is ultrastatic (i.e. $g_{00}=-1)$, then $g=-\det(g_{\mu\nu}),\,
R_{\,0}^{\,0}=0$, and
 (\ref{e6}) takes the form
\begin{equation}
 U_{\sc}(x)= \frac{1}{2}\, q^2\,G(x,x)\, ,\nn
  \end{equation} where $G$ is the
solution of the equation
\begin{equation}
 \partial_{\mu}\left(\sqrt{\mathfrak{g}}\, g^{\kn \mu\nu}\hsp\partial_ {\nu} G(x, x')\right)
 -\xi\, \mathfrak{R}\, \sqrt{\mathfrak{g}} \,G(x, x')
  =- \sqrt{\mathfrak{g}}\,\delta^{\kn d-1}(x, x')\,
\end{equation}
where $\mathfrak{g}=\det(g_{\mu\nu})$ and $\mathfrak{R}$ is the
corresponding scalar curvature. That is, $G$ is the Green's
function on the \mbox{$(d-1)$}\kn-dimensional space with the
metric $g_{\mu\nu}$  with Euclidean signature and the curvature $\mathfrak{R}$.

Now let us suppose that there exists at least one flat
extra spatial dimension, say $x_{d-1}$. Then formally identifying
$ix_{d-1}=t$, one notices that  the equation (\ref{e5}) for the
static
 Green's function  coincides with the full field equation (\ref{oper_1b}) for the \textit{Euclidean}
 Green's function $G^E(x,x'\mi d-1,n)$ in the spacetime with $(d-1)$ spacetime dimensions and $n-$dimensional conical
subspace. Finally, with the use of well-known relation between
 Euclidean $G^E$ and Feynman $G^F$ Green's  functions, we obtain,
 that in this case
\begin{equation}\label{e6qq}
U_{\sc}= \frac{1}{2}\, q^2\,G^{E}(x,x\mi d-1,n)=-\frac{i}{2}\,
q^2\,G^{F}(x,x\mi d-1,n)\, , \qquad\qquad {\mathbf{F}}=-
 \left(\frac{r}{r_0}\right)^{\!\bs}\frac{\delta U}{\delta \mathbf{r}}\,.
\end{equation}

One can study the self-energy of a static electric charge  along the
same lines.

In this case the solution of the Maxwell equations is static, with
the $d-$potential $A_M=(A_0(x), 0, \pp, 0)$ if the current equals
$J^M=(J(x), 0, \pp, 0))$. The only nontrivial component of
the Maxwell equations is
\begin{equation}\label{espot}
\partial_{\mu}\left(\sqrt{- g}~g^{00}g^{\mu\nu}
\partial_{\nu} A_0\right)=-\sqrt{-g}~J\,,
\end{equation}
and for the electrostatic self-energy one obtains (e.g. see
\cite{Smith})
\begin{align}
U_{\em}=&-\frac{1}{2}\int d^{d-1}x~ \sqrt{- g}~A_0 J =\nn \\ =  &
\frac{1}{2} \int d^{d-1}x~ \sqrt{- g}\int d^{d-1} x' \sqrt{- g'}
J (x)\kn G(x, x')\kn J(x')\,,
\end{align}
where Green's function of the Eq. (\ref{espot}) is defined as the
solution of
\begin{align}\label{esgf}
&\!\partial_{\mu}\left(\sqrt{- g}~g^{00}g^{\mu\nu}\partial_ {\nu}
G(x, x')\right)=\sqrt{- g}~\delta^{d-1}(x, x')\,.
\end{align}

So, for the point charge, when the charge density
$J=e\,\delta^{d-1}(x, x')$, we obtain
\begin{equation}
U_{\em} =\frac{1}{2}\, e^2\,G(x,  x) \, .
\end{equation}

In the particular case of an ultrastatic space Eq. (\ref{esgf})
takes the form
\begin{align}
\partial_{\mu}\bigl(\sqrt{\mathfrak{g}}~g^{\mu\nu}\partial_
{\nu} G(x, x')\bigr)=-\sqrt{\mathfrak{g}}~\delta^{d-1}(x, x')\,.
\nn
\end{align}
This equation coincides with Eq. (\ref{e6}) if $\xi=0$. Using this
fact one finds that
\begin{equation}\label{emse}
U_{\em}= \frac{1}{2}\, e^2\,G^{E}(x,x\mi d-1,n)\Bigl|_{\xi=0}\, .
\end{equation}

Consequently, on the background under consideration the electrostatic
self-energy can be obtained from the scalar one if we put $\xi=0$
and replace $q^2$ by $e^2$.


The spacetime of interest here (\ref{Metnew}), i.e. $d$-dimensional
spacetime with $n$-dimensional subspace with a solid or planar
angle deficit, satisfies the ultrastaticity condition, so we will use simple formulae  (\ref{e6qq}, \ref{emse}) for it.

\section{Green's function: perturbation theory}\label{PertTheo}

For our background metric (\ref{Metnew})
the exact
Green's function is unknown. Taking into account the fact that
\mbox{$(1-\beta) \ll 1$} we make use of the perturbation-theory
techniques. The Feynman propagator for the scalar field in curved
background satisfies the equation\footnote{In what follows we
define the Feynman propagator as $G^F(x, x')=i \langle
T\left[\phi\kn(x)\, \kn\phi\kn(x')\right]\rangle_{\rm vac}$. }
\begin{align}\label{oper_1b}
 {\cal{L}}(x, \partial)\,G^F(x,x' \mi d,n)= - \delta^{\smhsp d}\nhsp \smhsp
 (x-x')\,,
\end{align}
where ${\cal{L}}(x, \partial)$ stands for the field-equation operator and  determined by the background metric.

Following Schwinger \cite{Schwinger}, we rewrite eq.\,(\ref{oper_1b}), in the operator form
\begin{align}\label{eq1} \mathfrak{L}\smhsp \mathcal{G}= -1\,,  \qquad \qquad
\mathcal{G}= -\mathfrak{L}^{-1}\,.
\end{align}
If operator $\mathfrak{L}$ allows to be expressed as $
\mathfrak{L}=\mathfrak{L}_0 +\delta\mathfrak{L}$, where
$\delta\mathfrak{L}$ is considered as a small perturbation, then
representing the solution of eq.\,(\ref{eq1}) in the form $
\mathcal{G}=\mathcal{G}_0+\delta\mathcal{G}$, with
$\mathcal{G}_0=-{\mathfrak{L}_0}^{-1}$ being the unperturbed
Green's function, one obtains
\begin{align}\label{oper7}
\mathcal{G} & = \left[-\mathfrak{L}_0 \left(1-\mathcal{G}_0
\delta\mathfrak{L}\right)\vp\right]^{-1} = \mathcal{G}_0+
\mathcal{G}_0 \hsp\delta \mathfrak{L}\hsp \mathcal{G}_0 +
\mathcal{G}_0 \hsp\delta \mathfrak{L}\hsp \mathcal{G}_0\hsp\delta
\mathfrak{L}\hsp \mathcal{G}_0 +\pp\,.
\end{align}

In the case under consideration  $\mathfrak{L}_0$ is determined by
the zeroth order in the small quantity $(1-\beta)$, hence
 $$ {\cal{L}}_0(x, \partial)=
\partial^{\kn 2}\,,\qquad \partial^{\kn 2}\equiv\eta^{MN}\partial_M\partial_N\, .$$
The perturbation operator
\begin{equation}\label{dLem}
\delta{\cal{L}}(x, \partial)=\partial
_{M}\left(\sqrt{{-g}}\,\,g^{MN}\partial _{N}\right) -\partial^2 -
\sqrt{-g}\, \xi R\,
\end{equation}
to the first order in $(1-\beta)$ reads:
\begin{align}\label{e16's}
\delta{\cal{L}}(x,
\partial)=n\,\alpha(r)\biggl(\partial_0^2-\sum_{N=n+1}^{d-1} \!\!\partial_N^2\biggr) -
(n-2)\sum_{i=1}^{n} \Bigl[\alpha(r)\,\partial_i^2 + \bigl(\partial_i
\alpha(r)\bigr)\kn\partial_i\Bigr] - \xi \gamma(r)\, .
\end{align}

In order to compactify our equations below, let us introduce the notation
$$ \beta'=1-\beta\, .$$
With the use of this notation
$$\alpha(r)=\bs\,\ln r$$ and
\begin{align}\label{curvlin}
\gamma(r)=
\left\{
  \begin{array}{ll}
     4\pi \bs\kn\delta^2(\mathbf{r}), & \hbox{$n=2$\kn;} \\
     2(n-1)(n-2)\bs/{{r}}^2, & \hbox{$n \geqslant 3$.}
  \end{array}
\right.
\end{align}
with $\mathbf{r}=(x_1, x_2, \pp, x_n)$.

In the problem-at-hand the function $G_0^F(x,
x')=\bra{x}\mathcal{G}_0\ket{x'}=- \bra{x}\partial^{-2}\ket{x'}$
in Fourier basis takes the form\footnote{Hereafter the direct
Fourier-transform is defined as
\begin{align}
 \mathcal{F}[\varphi\kn(x)](q)= \int d^d x\, \varphi
\kn(x) \,\e^{-i qx}\,. \nn
\end{align}}:
\begin{align*}
G_{0}^F (x-x')=\int\frac{\ddp}{(2\pi)^d}\,
\frac{\e^{ip\,(x-x')}}{p^{\kn 2}-i\varepsilon}\, ,
\end{align*}
where $p^2={\mathbf{p}}^2-(p^0)^2$ and
$p\,x=\mathbf{p}\mathbf{x}-p^{0}x^{0}$.

For the first-order correction to the Green's function from (\ref{oper7}) we get the following expression:
\begin{align}\label{oper_8b}
 G^F_1(x, x'\mi d,n)& =    \bra{x}
\mathcal{G}_0 \hsp \delta\mathfrak{L}\hsp \mathcal{G}_0\ket{x'} =
\int \frac{\ddq}{(2\pi)^{d}}\,\e^{iqx}\,\int
\frac{\ddp}{(2\pi)^{d}}\,\e^{ip(x-x')}\,
 \frac{\delta {\cal{L}}(q,
ip) }{\left[p^{\kn 2}-i\varepsilon\right]\left[(p+q)^{\kn
2}-i\varepsilon\right]}\, ,
\end{align}
where $\delta {\cal{L}}^{(1)}(q, ip)$  is defined as:
\begin{equation}\label{e14gen}
\delta {\cal{L}}(q, ip)=\int \ddx\, \e^{-iqx}\!\left[\left. \delta
{\cal{L}}(x,
\partial)\right|_{\partial\to ip}\right] \, .
\end{equation}
Here one implies that the differential operator $\delta
{\cal{L}}(x,
\partial)$ is prepared to the form where all differential operators stand before (at right-hand side from)
the coordinate functions, and further one performs  the substitution
$\partial_j\to ip_j$ and calculates  the Fourier-transform,
considering $p_j$ as parameters.

In our problem the perturbation operator reads
(\ref{e16's}),
\begin{equation}\label{e14}
\delta {\cal{L}}(q,
ip)=\Bigl[np^2-2{\mathbf{p}}^2+(n-2)(\mathbf{q}\mathbf{p})\Bigr]{\cal{F}}[\alpha](q)-\xi
{\cal{F}}[\gamma](q)\ ,
\end{equation}
where $\mathbf{p}=(p_1, \pp, p_n)$ and $\mathbf{q}=(q_1,
\pp, q_n)$ are $n$\kn-dimensional conformal vectors  with
the Euclidean scalar product
$(\mathbf{q}\mathbf{p}) \equiv  \delta_{ik}q^ip^k$, while
$p^{\kn2} \equiv \eta_{MN}p^Mp^N$.

 Making use of the explicit form of operator
 $\delta {\cal{L}}(x, \partial)$,
substitution of (\ref{e14}) into eq.\,(\ref{oper_8b}) yields
\begin{align}\label{Obama}
G^F(x, x'\mi d, n)=G_0^F(x-x')\akn + \akn\akn \int
\akn\akn\frac{\ddq}{(2\pi)^d}\, \e^{iqx} \akn & \int\akn\akn
\frac{\ddp}{(2\pi)^d}\,\frac{\e^{ip(x-x')}}{\left[p^{\kn
2}-i\varepsilon\right] \left[(p+q)^{\kn
2}-i\varepsilon\right]}\times \nn\\ &  \times\Bigl[\Bigl(np^2\akn
-2{\mathbf{p}}^2+(n-2)(\mathbf{q}\mathbf{p})\Bigr){\cal{F}}[\alpha](q)-\xi
{\cal{F}}[\gamma](q)\Bigr]\, .
\end{align}

Taking into account that formulae for the background curvature
(\ref{curvlin}) differ for cases \mbox{$n=2$} and
\mbox{$n\geqslant3$}, we consider here the generic case of a
global monopole, while the case of a cosmic string is delegated to
the Section\,\ref{Cosmostring} below.


In this case (\ref{Obama}) takes the form
\begin{align}\label{Obama2}
 G^F(x, x'\mi d, n)= G_0^F(x-x')-\frac{\Gamma(n/2)}{2\,\pi^{n/2}} & \int d^{\kn n} q\,
\frac{\e^{i\mathbf{q}\mathbf{x}}}{({\mathbf{q}}^2)^{n/2}}  \int
\frac{\ddp}{(2\pi)^d}\, \akn\akn \frac{\e^{ip(x-x')}}{\left[p^{\kn
2}-i\varepsilon\right] \left[(p+q)^{\kn
2}-i\varepsilon\right]}\times \nn\\ &
\times\left[{np^{\kn
2}-2\mathbf{p}}^2+(n-2)(\mathbf{q}\mathbf{p})+2\,\xi(n-1)\kn{\mathbf{q}}^2\right]
.\end{align}
 where we use the following well-defined  Fourier-transforms
\cite{Shilov}:
\begin{align}\label{Four1}
&\mathcal{F}\!\left[\ln r\right]\!(q)=-\frac{2^{d-1}}{\pi^{n/2-d}}
\frac{\Gamma(n/2)}{(\mathbf{q}^2)^{n/2}}\,\delta(q^0)\prod\limits_{N=n+1}^{d-1}\delta(q^N)\, ,\\
&\mathcal{F}\!\left[r^{-\lambda}\right]\!(q)=\frac{2^{d-\lambda}}
{\pi^{n/2-d}}\frac{\Gamma[(n-\lambda)/2]}{\Gamma[\lambda/2]}\frac{1}{(\mathbf{q}^2)^{(n-\lambda)/2}}\,
\delta(q^0)\prod\limits_{N=n+1}^{d-1}\delta(q^N)\,.
\end{align}

In Eq. (\ref{Obama2}) and in all subsequent equations $q \equiv
(0, \mathbf{q},\underbrace{0,0 \pp 0}_{d-n-1})\,.$

All the quantities we are interested in, are expressed via  the
Feynman propagator $G^F(x, x'\mi d, n)$  and its derivatives,
evaluated in coincident points. The corresponding expressions
diverge, and for their evaluation we make use of the
dimensional-regularization method
  (see, e.g.
\cite{Collins}).

The dimensional regularization consists in the replacement of the
determining  function $G(x, x)$ by $G_{\reg}(x, x)$, corresponding
formally to the Green's function in \mbox{$D=(d-2\varepsilon)$}
dimensions. The subsequent renormalization includes the splitting
of $G_{\reg}(x, x)$ onto two parts; the first one diverges as
$\varepsilon\rightarrow 0$, while the other is finite.  The
renormalization finishes with the neglect of the divergent part
$G_{\rm div}(x, x)$, with subsequent computation of the limit
$\varepsilon\rightarrow 0$. But as it was remarked by Hawking
\cite{Hawking}, in the case of a curved space this procedure may
be ambiguous, because in general there can be a variety of
different ways of performing the analytic continuation from $d$ to
$D$ dimensions. The simplest way is to take the product of the
initial $d$-dimensional spacetime with a flat space with $D-d$
dimensions with subsequent analytic continuation with respect to the
extra dimensions.

Fortunately, the spacetime of interest here, has  originally the
structure demanded by this prescription. So, according to
Hawking's prescription, we will define $ G^F_{\ren}(x, x\mi d, n)$
as a limit
\begin{equation}\label{n3}
 G^F_{\ren}(x, x\mi d, n)=\lim\limits_{\varepsilon\to
0}\left[G^F_{\reg}(x, x\mi D, n) - G^F_{\rm div}(x, x\mi D,
n)\vp\right].
\end{equation}

As it was shown by Hawking,  results obtained by this
prescription  are in agreements with those ones obtained with help
of the method of generalized $\zeta-$function.

\section{Renormalized $  \boldsymbol{\langle}
\boldsymbol{\phi^{\kn 2}(x)}\boldsymbol{\rangle}$}\label{phi2ren}

Now proceed to a perturbative expression  for the regularized
value of vacuum averaged $\langle \phi^{\kn 2}(x)\rangle_{\ren}$:

We define the Feynman propagator as $G^F(x, x')=i \langle \kn
T\left[\phi\kn(x)\, \kn\phi\kn(x')\right]\rangle_{\rm vac}$\kn. So,
\begin{equation}
\langle\varphi^2(x)\rangle_{\ren}=-i\,G^F_{\ren}(x, x\mi  d,
n)=G^E_{\ren}(x, x\mi  d, n)\,.
\end{equation}

The first problem, arising here, is an expression arising in the
zeroth order in $\bs$. Indeed, for the contribution from the
first term on the right hand side of (\ref{Obama2}) to the Green's
function  taken in the limit of coincidence points, we have the
formally divergent expression
$$G_0^F(x, x)= -\frac{1}{(2\pi)^d}\,\int\frac{\ddp}{p^{\kn 2}}\, .$$

However, all integrals of the form
\begin{equation}\label{tadp}
\int \ddp \,\frac{p_{i_1}\pp\kn p_{i_k}}{p^{\kn 2}}\, ,
\end{equation}
which diverge in UV- or/and in IR-limits and correspond to the
<<tadpole>>-type diagrams in QFT, are set to have zero value (no tadpole
prescription)
within  the dimensional-regularization technique  (see, e.g. \cite{Smirnov}). According to this
prescription   we shall put all terms of the form
(\ref{tadp}) equal to zero.

Thus,  in the case $d\geqslant 4\, , 3\leqslant n\leqslant d-1$
and arbitrary value of the coupling constant $\xi$,
 for the first non-vanishing contribution to the
coincidence-points Green's function one obtains from eq.\,(\ref{Obama2}):
\begin{align}\label{g1_4}
 & G^F(x, x\mi d, n)=\bs\,\frac{\Gamma(n/2)}{2\,\pi^{n/2}} \int d^{\kn n} q\,
\frac{\e^{i\mathbf{q}\mathbf{x}}}{({\mathbf{q}}^2)^{n/2}}  \int
\frac{\ddp}{(2\pi)^d}\,
\frac{2\kn{\mathbf{p}}^2-(n-2)(\mathbf{q}\mathbf{p})-2\,\xi(n-1)\kn{\mathbf{q}}^2}{\left[p^{\kn
2}-i\varepsilon\right] \left[(p+q)^{\kn 2}-i\varepsilon\right]}\,
.\end{align}

The integral over $d^d p$ diverges. However, it has a standard
form for the QFT. Within the framework of the dimensional
regularization one performs the Wick rotation
$$ p^0\rightarrow ip^{\kn 0}_E\, ,\quad \ddp\rightarrow id^d p_E\, $$
and  replaces the integral over $d^d p_E$   by the expression that
formally corresponds to integration over a
\mbox{$(D-2\varepsilon)$}\kn-dimensional $p_E$-space:
\begin{align}\label{regul}
\int\frac{\ddp}{(2\pi)^{\kn d}} \,\pp \rightarrow
i\,\mu^{2\varepsilon}\int\frac{\dDp_E}{(2\pi)^D }\, \pp \, .
\end{align}
An arbitrary parameter $\mu$ with the dimension of reciprocal length
is introduced to preserve the dimensionality of the regularized
expression.

Computational technique for these integrals is well-developed
(e.g., see \cite{Smirnov}) and we obtain\footnote{For brief
reference, we overview derivation of some of them in the
Appendix\,\ref{App_ints}. }:
\begin{align}\label{g8}
& \int\frac{\dDp_E}{(2\pi)^D}\, \frac{2\kn
{\mathbf{p}}^2-(n-2)(\mathbf{q}\mathbf{p})-2\kn
\xi(n-1)\kn{\mathbf{q}}^2}{p_{E}^{\kn 2}\,(p+q)_E^2}
=\left(1-\frac{\xi}{\xi_D}\right) \frac{2(n-1)}{(4\pi)^{D/2}}
\frac{\Gamma^2 \akn(D/2)}{\Gamma(D)}\,\frac{\Gamma(2-D/2)}
{({\mathbf{q}}^2)^{1-D/2}}\, ,
\end{align}
where we have denoted
\begin{align}
\xi_D \equiv  \frac{D-2}{4\kn(D-1)}\,.\nn
\end{align}
Notice, when \mbox{$\varepsilon=0$} and $\xi_D=\xi_d$ the field
equation for a massless  scalar field $\phi$ is invariant under
conformal transformations of the metric.

For even $d$ the expression  (\ref{g8}) has a simple pole at
$\varepsilon=0$, and under the removal of regularization the divergence in
\mbox{$G^E_{\reg}(x, x \mi D, n)$} may arise due to this pole, or due to the   $d^{\kn n} q$-integration, or due to the both reasons simultaneously.

Let consider this question in more details.

Substituting (\ref{g8})  into (\ref{g1_4}) and making use of the integral (\ref{Four1})
for the regularized vacuum mean $\langle\phi^{\kn 2}(x)\rangle$ we obtain (for all $3\leqslant n \leqslant (d-1)$) the following expression:
\begin{align}\label{g4}
 \langle\phi^{\kn 2}(x) \rangle_{\reg}=-iG^F_{\reg}(x, x\mi D, n)=
 \mu^{2\varepsilon} \bs
\frac{n-1}{4\kn \pi^{D/2}}
\frac{\Gamma(n/2)\,\Gamma^3\akn(D/2)}{\Gamma(D)}\Bigl(\frac{\xi}{\xi_D}-1\Bigr)\,\frac{\Gamma\!\left(-\frac{D-2}{2}\right)}{\Gamma
\bigl(-\frac{D-n-2}{2}\bigr)}\,\frac{1}{r^{D-2}}\,.
\end{align}

We see that the behavior of the regularized VEV $\langle\phi^{\kn
2}(x)\rangle_{\reg}$ in the limit \mbox{$\varepsilon\rightarrow
0$} is determined by the factor
\begin{align}\label{gamma}
 { \Gamma\akn \Bigl(-\frac{D-2}{2}\Bigr)}\Bigr/{\Gamma
\akn \Bigl(-\frac{D-n-2}{2}\Bigr)}
\end{align} and, therefore,
depends significantly upon the parity of the dimensionality both of the entire $d$-dimensional bulk and of its $n$-dimensional
conical subspace.

Let consider all possible  cases.

\medskip

\noindent $\bullet$\;\,\textbf{even $\boldsymbol{d}$,
 odd $\boldsymbol{n}$.} In this case \mbox{$(d-n-2)/{2}$}
is  semi-integer, so Gamma-function
in denominator (\ref{gamma}) takes its finite and nonzero value. Whereas the Gamma-function
 \mbox{$\Gamma\akn\left(1-D/2\right)$} in the numerator of eq.\,(\ref{gamma}) has a simple pole in \mbox{$\varepsilon=0$}, thus
when the regularization removed, the separation of divergent part may be performed
with help of the Laurent expansion
\begin{align}\label{resi}
\Gamma(-m+\varepsilon)=\frac{(-1)^{\kn m}}{m!}\left(\frac{1}{\varepsilon}-\gamma
 + H_m  + {\cal{O}}(\varepsilon)\right),
\end{align}
where~$\gamma$  is the Euler's constant, and  \mbox{$
H_m=\sum\limits_{k=1}^m k^{-1}$} is the $m$-th harmonic number.

We obtain now
\begin{align}
 &\langle\phi^{\kn 2}(x)\rangle_{\rm div}= -i G^F_{\rm div}(x, x\mi  d, n)=\frac{(-1)^{\kn d/2}}{\varepsilon}
  \frac{\bs}{2\kn\pi^{d/2}}   \frac{(n-1)}{(d-n)}\, \frac{\Gamma(n/2)\,\Gamma^2(d/2)}{ \Gamma(d)\,\Gamma
\bigl(-\frac{d-n}{2}\bigr)} \left(\frac{\xi}{\xi_d}-1\right)
\frac{1}{r^{d-2}}\, .\nn
\end{align}
Notice, in the case of a conformal coupling \mbox{$\langle\phi^{\kn 2}(x)\rangle_{\rm div}$} vanishes.

Separation of the finite part of the regularized expression
(\ref{g4}) is achieved by the following expansions:
$$ \frac{\xi}{\xi_D}-1=\left(\frac{\xi}{\xi_d}-1\right)+
\varepsilon\,\frac{8\kn\xi}{(d-2)^2}+\mathcal{O}(\varepsilon^2)\,, \qquad\qquad    \frac{f(D)\,\mu^{2\varepsilon}}{r^{D-2} }=\frac{f(d)}{r^{d-2} }
\left[1+2\kn\varepsilon\!\left(\ln \mu r-\frac{f'(d) }{f(d)} \right)
+ \mathcal{O}(\varepsilon^2) \right], $$ where
$$ f(z) \equiv
\frac{ \Gamma^3(z/2)}{ \pi^{z/2}\Gamma(z)\,\Gamma
\bigl(\frac{2+n-z}{2}\bigr)} \, ,$$ that leads to the final result
\begin{align}\label{g4ren_a1}
 \langle \phi^{\kn 2}(x)\rangle_{\ren} =   (-1)^{(n-1)/2}\kn
\bs \frac{(n-1)\,\Gamma(n/2)\,\Gamma\!\left(\frac{d-n}{2} \right)}{2
\pi^{d/2+1}}
\frac{\Gamma^2(d/2)}{\Gamma(d)}\left[
\left(\frac{\xi}{\xi_d}-1\right)  \ln \tilde{\mu} r +\frac{1
}{(d-1)(d-2)} \right]\,\frac{1}{r^{d-2}}\, .
\end{align}
The constant $\tilde{\mu}$ here is a renormalized value of the
constant  $\mu$ introduced above:
$$\tilde{\mu}=\mu\exp\!\left(-\frac{f'(d) }{f(d)} +  \frac{
   H_{d/2-1}-\gamma}{2}+\frac{1}{(d-1)(d-2)}\right).$$

Notice, with the conformal coupling the logarithmic term and the uncertainty related with the arbitrary constant $\tilde{\mu}$ in it,
 disappear from $\langle \phi^{\kn 2}(x)\rangle_{\ren}$.

Separately, we consider the case of higher-dimensional monopole, where $n=(d-1)$.
Then from  eq.\,(\ref{g4ren_a1}), making use of well-known formulae on Gamma-function
\begin{align}\label{gammae}
\Gamma(x)\,\Gamma(1-x)=\frac{\pi}{\sin \pi x}\,,\qquad\qquad  \Gamma(2x)=\frac{2^{\kn 2x-1}\Gamma(x)\,\Gamma(x+1/2)}{\sqrt{\pi}}\,,
\end{align}
we have:
\begin{align}\label{g4ren_b}
& \langle \phi^{\kn 2}(x)\rangle_{\ren} =
  (-1)^{d/2-1}
\frac{\bs\,(d-2)\,\Gamma(d/2)}{2^{ d-1}\pi^{ d/2}(d-1)} \left[ \left( \frac{\xi}{\xi_d}-1\right) \ln \tilde{\mu} r
+\frac{1 }{(d-1)(d-2)}  \right]\frac{1}{r^{d-2}}\,.
\end{align}
In particular, for the spacetime types $(4, 3)$ and $(6, 5)$ one obtains:
\begin{align}\label{g4ren_4}
 &\langle \phi^{\kn 2}(x)\rangle_{\ren} =
 -\frac{\bs}{12\,\pi^{2}r^{2}}
\left[ \bigl( 6\kn\xi-1 \bigr)  \ln \tilde{\mu} r  +\frac{1 }{6}  \right], \quad d=4;  \\
&\langle \phi^{\kn 2}(x)\rangle_{\ren} =
\frac{\bs}{20\,\pi^{3}r^{4} } \left[ \Bigl(5\kn\xi-1\Bigr)
\ln \tilde{\mu} r  +\frac{1 }{20} \right],\quad d=6,  \label{g4ren_6}
\end{align}
that coincides (with the accuracy required) with the results of
\cite{Mazzitelli} and \cite{Bezerra_2002}, respectively\footnote{A
note to be added: our  result (\ref{g4ren_4}) coincides with that
one of \cite{Mazzitelli} numerically, since in the cited work it
is the numerical
 computation that was done for several introduced (within their computational scheme)
 integrals (namely, \cite[eqns.(2.18, 2.19)]{Mazzitelli}). However, these integrals may be computed
  analytically; doing this, the result coincides with our (to the leading in $\bs$ order we are interested here).}.

Finally, the renormalized  VEV of
$\langle \phi^{\kn2}(x)\rangle$ for 3-dim monopole in six dimensions reads
 \begin{align}\label{Tmn_ren6D3n}
& \langle \phi^2(x)\rangle_{\ren} =- \frac{ \bs }{
120\kn\pi^{3}r^4} \left[ \Bigl(5\xi-1\Bigr)  \ln \tilde{\mu} r +\frac{1
}{20} \right],  \qquad d=6,n=3.
\end{align}

\smallskip

\noindent $\bullet$\;\,\textbf{odd $\boldsymbol{d}$  and
$\boldsymbol{n}$.} Here as \mbox{$\varepsilon\rightarrow 0$} the
Gamma-function \mbox{$\Gamma\akn\left(1-D/2\right)$} in numerator
is finite, while \mbox{$\Gamma\akn\left(1-(D-n)/2\right)$} in the
denominator of  (\ref{g4}) has a simple pole. Therefore, in the
lowest in $\bs$ order  $\langle \phi^{\kn 2} (x) \rangle_{\ren}\,$
vanishes\kn\footnote{\label{footinootie}Unfortunately, we may say
nothing about
 the result in the second order: whether it also vanishes, or has the finite value.
 Probably, the non-perturbative approach in some particular case $(d,n)$ of this
 type may shed light on this problem. Investigation of these effects lies beyond
 the mainline of our work here and hopefully will be considered later.}.

\medskip

\noindent $\bullet$\;\textbf{odd  $\boldsymbol{d}$, even
$\boldsymbol{n}$.} In this case the both  Gamma-functions,
$\Gamma\akn\left(1-D/2\right)$  and
$\Gamma\akn\left(1-(D-n)/2\right)$ in (\ref{gamma}), are finite,
hence $\langle \phi^{\kn 2}(x)\rangle_{\rm div}=0$ and after some
algebra (\ref{gammae}) we arrive at
\begin{align}\label{g4_www}
\langle \phi^{\kn 2}(x)\rangle_{\ren}=& (-1)^{\kn
n/2}\frac{\bs}{4\pi^{d/2}}
\frac{(n-1)\kn\Gamma(n/2)\,\Gamma^2(d/2)} {\Gamma(d)}\, \Gamma\akn\left(\frac{d-n}{2}\right)\!
\left(\frac{\xi}{\xi_d}-1\right) \frac{1}{r^{ d-2}}\, .
\end{align}
Hence for the $d-$dimensional monopole, $(d, d-1)-$spacetime, we
have
\begin{align}\label{g4renuuu}
\langle \phi^{\kn 2}(x)\rangle_{\ren}= (-1)^{ n/2}\bs
 \frac{(d-2) \kn \Gamma(d/2)}{2^{d} \pi^{ d/2-1}(d-1)\,r^{ d-2}}\left(\frac{\xi}{\xi_d}-1\right) .
\end{align}
In particular, for the five-dimensional monopole \mbox{$(d=5$}, \mbox{$n=4)$}
VEV $\langle \phi^{\kn 2}(x)\rangle_{\ren}$ takes the form
\begin{align}\label{g4ren_5}
\langle \phi^{\kn 2}(x)\rangle_{\ren}=  \bs
  \frac{3\kn( 16 \,\xi-3 )}{2^9 \pi r^{3}}\,,
\end{align}
which coincides with the results of the papers
\cite{Bezerra_2002,Bezerra_2006}.

 \medskip

\noindent $\bullet$\;\,\textbf{even $\boldsymbol{d}$  and
$\boldsymbol{n}$.} Here the simple pole of
$\Gamma(1-D/2)$ in numerator  (\ref{gamma}) is compensated by that one of the
Gamma-function $\ds\Gamma\bigl(-(D-n)/2\bigr)$ in denominator.  The result of
(\ref{gamma}) at \mbox{$\varepsilon=0$}, thereby, equals the ratio of the corresponding residuals. Moreover, as in the previous case, the divergent part
\mbox{$\langle \phi^{\kn 2}(x)\rangle_{\rm div}$} vanishes, and VEV equals
\begin{align}\label{g4_even}
\langle \phi^{\kn 2}(x)\rangle_{\ren}= &(-1)^{n/2}
 \bs \frac{(n-1)\,\Gamma\!\left(\frac{d-n}{2}
\right)}{4\kn \pi^{ d/2}}
\frac{\Gamma^2(d/2)\,\Gamma(n/2)}{\Gamma(d)}
\left(\frac{\xi}{\xi_d}-1\right) \frac{1}{r^{d-2}}\,,
\end{align}
and thus $\langle \phi^{\kn 2}(x)\rangle_{\ren}$ vanishes (in
lowest in $\bs$ order) for the case of the conformal scalar field.

A direct comparison of (\ref{g4_www}) with the formula (\ref{g4_even})
shows, that in the interested accuracy the two cases with even
conical subdimensionality  $n$ can be combined into the unified one,  despite the intermediate formulae were based on the drastically different behavior of the Gamma-function. However, for   odd
 $n$ the result depends significantly upon the parity of the bulk's dimensionality.

Summarizing, in this section we have computed the renormalized
vacuum averaged  $\langle\phi^{\kn 2}\rangle_{\ren}$ for a
massless scalar field on the generalized background
(\ref{Metnew}). We have made the computation up to the first order
in $\bs$ but for arbitrary values of the coupling constant $\xi$
and for any dimension of the space \mbox{$d\geqslant 4$} and any
dimension of its conical subspace in the interval
\mbox{$3\leqslant n\leqslant d-1$}. For doing so we have used
perturbation technique combined with the method of dimensional
regularization. For the  case with even $d$ and odd $n$ (in
particular, for the four dimensional global monopole) it is the
logarithmic factor $\ln \mu r$ that has the crucial significance
for the field with nonconformal factor, since all finite
non-logarithmic terms may be absorbed by the finite
renormalization of $\mu$.

 The
methods presented in this section, may be used to compute the
renormalized mean value of the energy-momentum tensor in a similar way.

\section{Renormalized energy-momentum tensor}\label{EMTVEV}

The total energy-momentum tensor derived from the action (\ref{action1}), is given by  (\ref{EMT1}). In terms of the Green's function, the regularized VEV of the energy-momentum tensor is given by
\begin{align}  \label{putinmerde}
\langle T_{MN}(x) \rangle_{\rm reg}= -i\lim_{x'\rightarrow x}\
D_{MN}  \, G_{\rm reg}^F (x,x')\,,
\end{align}
where $D_{MN}$ stands for the appropriate differential operator ($\nabla^{M} $  and $\nabla^{M'} $ denote the covariant derivative over $x^{M}$ and $x'^{M}$, respectively):
\begin{align}
D_{MN}=(1-2\xi) \nabla_{ \akn M}\nabla_{  N'} \!+
\frac{1}{2}\,(4\xi-1)\, \nabla_{ \akn L}\nabla^{
L'}\,g_{MN} + \xi\,\Bigl[ R_{MN} -\frac{1}{2}\,R
\,g_{MN}+2\kn \nabla_{ \akn L} \nabla^{ L}
\,g_{MN} -2 \kn \nabla_{\akn M} \nabla_{\akn N}\Bigr]\, .\nn
\end{align}

Taking into account the special significance of the minimally coupled field, and in order to dilute routine computations,  it is natural to compute the renormalized vacuum momentum density separately for different powers of $\xi$.  We start to separate $\xi-$terms already from definition:
thereby we can split energy-momentum tensor  as
\begin{align}
T_{MN}=T_{MN}^{(0)}+\xi T_{MN}^{(\xi)} \nn
\end{align}
with
\begin{align}\label{Tmndxi}
& T_{MN}^{(0)} =\phi_{, \kn M} \phi_{, \kn N}-\frac{1}{2}\,g_{MN} \, \phi_{,\kn L} \phi^{\, , \kn L}\nn \\
& T_{MN}^{(\xi)}=  - 2  \kn \phi_{,\kn M} \phi_{,\kn N} + 2 \kn  \phi_{,\kn L} \phi^{\,, \kn  L}  \,g_{MN}-  2 \kn \phi_{\kn ;\kn  M N}\kn \phi   +
 2\kn\phi\; \Box\phi\,g_{MN}+\frac{1}{2}\,\bigl(2 R_{MN} -R \,g_{MN}\bigr)\, \phi^{\kn 2}\,.
\end{align}
Each term here contains a quadratic form on $\phi\,$ and, therefore, can be derived from the Feynman propagator. Hence we may apply our point-splitting procedure for the derivatives combined with the perturbation-theory scheme, to reveal the linear on $\bs $ contributions.

A note to be mentioned: $T_{MN}^{(\xi)}$ contains the second  covariant derivatives; computing them, one needs in the corresponding Christoffel symbols.
 In the coordinates specified, 
all non-vanishing Christoffel symbols are of order $\mathcal{O}\kn(\bs )$. Given that the zeroth (in  $\bs $) order of
the  Green's function vanishes in our scheme (as no tadpole prescription), the retaining of the Christoffel-part contribution
yields the order $\mathcal{O}\kn({\bs }^2)$, i.e. exceeds the necessary accuracy. Hence we can neglect these terms and consider derivatives as <<flat>>.

Repeating the steps to construct the Green's function, the 1st-order operator correction $\delta{\cal{L}}(x,
\partial)$  also can be split as $\delta{\cal{L}}(x,
\partial)= \delta{\cal{L}}^{(0)}(x,
\partial) +\xi \delta{\cal{L}}^{(\xi)}(x,
\partial)$ with\kn\footnote{Within  this section the index $\sigma$ runs over all <<flat>> indices: $\sigma=0, n+1, \pp, d-1$.}
\begin{align}\label{e16'}
&\delta{\cal{L}}^{(0)}(x,
\partial)=-n\alpha(r)  \kn \partial_{\sigma}\kn \partial^{\sigma} -
(n-2)  \Bigl[\alpha(r)\kn \partial_i \partial^i + \bigl(\partial_i
\alpha(r)\bigr)\partial^i\Bigr]\,, \nn \\
 & \delta{\cal{L}}^{(\xi)}(x,
\partial)= -   R(r)\,.
\end{align}
In what follows, the energy-momentum VEV in the first non-vanishing order reads  schematically:
\begin{align}
T_{MN}={}^{0}T_{MN}+{}^{1}T_{MN} \,\xi+{}^{2} T_{MN} \xi^2,
 \end{align}
where
\begin{align}\label{T012}
&{}^{0}T_{MN}= T_{MN}^{(0)}\bigl[\delta{\cal{L}}^{(0)}\bigr]\nn\\
& {}^{1}T_{MN}= T_{MN}^{(0)}\bigl[\delta{\cal{L}}^{(\xi)}\bigr] +T_{MN}^{(\xi)}\bigl[\delta{\cal{L}}^{(0)}\bigr] \nn \\
&{}^{2} T_{MN}=T_{MN}^{(\xi)}\bigl[\delta{\cal{L}}^{(\xi)}\bigr].
\end{align}
The non-vanishing components of the Ricci tensor in our coordinates  are given by (\ref{Riccies}) and survive in the conical sector only.
By this reason, we should neglect the curvature-term in the last term  in (\ref{Tmndxi}) since it contributes as $\mathcal{O}\kn(\bs^2)$. Furthermore, after the replacement of the covariant derivatives by simple ones, the d'Alembert operator in   $g_{MN}\phi \,\Box\phi$ adds the multiplier $p^{\kn2}$ into the numerator of the Fourier integral. Multiplying by $p^{-2}(p+q)^{-2}$, this leads to the single-propagator Fourier integral, which vanishes in our scheme. Thereby, we can neglect this term also and  replace  (\ref{Tmndxi}) by its effective expression:
\begin{align}\label{Tmndxi_eff}
T_{MN}^{(\xi)}=  - 2  \phi_{, M} \,\phi_{, N} +  2  \phi_{, L} \phi^{\,, L} \,g_{MN}- 2  \phi_{, M N}\,\phi   \,.
\end{align}
Performing the Fourier-transforms in (\ref{e16'}), the $\xi-$separation in $\delta {\cal{L}}(q,ip)$ reads effectively
\begin{align}\label{xi0}
&\delta {\cal{L}}^{(0)}(q,ip)=-2^{n-1}\pi^{n/2}(2\pi)^{D-n}\Gamma(n/2)\,\delta^{D-n}(q^{\sigma}) \frac{-2\kn{\mathbf{p}}^2+(n-2)\mathbf{q}\mathbf{p}}{|\mathbf{q}|^{n}} \\
\label{dxi}
&\delta {\cal{L}}^{(\xi)}(q,ip)=-2^{n}(n-1)\,\pi^{n/2}(2\pi)^{D-n}\Gamma(n/2)\,\delta^{D-n}(q^{\sigma }) |\mathbf{q}|^{-(n-2)},
\end{align}
so  the latter actually does not depend upon $p^{M}$.

\subsection{Computation of $\boldsymbol{\langle T_{MN}\rangle_{\rm ren}}$ with minimal coupling}

Starting from (\ref{xi0}) and
proceeding along the same lines as for $\langle\phi^{\kn 2}\rangle$, we
obtain:
\begin{align}
 \langle {\kn}^{0} T_{MN}(x)\rangle_{\reg}=\bs\int \frac{d^D q \, d^D
p}{(2\pi)^{2D}}\,\e^{iqx}  \frac{\delta {\cal{L}}^{(0)}(q,ip) }{\left[p^{\kn2}-i\varepsilon\right]\left[(p+q)^2-i\varepsilon\right]} \Bigl(p_M p_N+ q_M  p_N -\frac{1}{2}\, \eta_{MN}\, \mathbf{q} \mathbf{p}  \Bigr).
\end{align}
After the integration with help of integrals of Appendix \ref{App_ints}, $\langle {\kn}^{0}T_{MN}(x)\rangle_{\reg}$ reads
\begin{align}
\langle {\kn}^{0}T_{MN}(x)\rangle_{\reg}=&- \bs\frac{ \Gamma(n/2)\,\Gamma^2(D/2)\,\Gamma\akn\left(-\frac{D-2}{2}\right)}{2^{D+3}\pi^{(D+n)/2}(D+1)\,\Gamma(D)}
 \int d^{\kn n}\akn \mathbf{q} \,\frac{\e^{i\mathbf{qx}}}{ |\mathbf{q}|^{2+n-D}} \times \nn
 \\& \times\Bigl[- A_D\, \tq_{M} \tq_{N}+(n-1)(D^2-2D-2) \,\mathbf{q}^2  \, \eta_{MN}+2\, \mathbf{q}^2 \,\te_{MN}\Bigr]
\end{align}
with $A_D \equiv D\kn(D-n)+(n-2)(D-2)(D+1)$. Hereafter the <<tilded>> quantity with indices means that it equals the corresponding tensor with no tilde for conical-subspace index, and vanishes in the opposite case.

Integrating the remaining Fouriers,  one arrives at
\begin{align}\label{Tmn_ren}
 \langle {\kn}^{0}T_{MN}(x)\rangle_{\reg}=&   \bs\frac{ \Gamma(n/2)\,\Gamma^3\akn(D/2)\,\Gamma\akn\left(-\frac{D-2}{2}\right)}{2^{3}\pi^{D/2}(D+1)\,\Gamma(D)\,\Gamma\akn\left(-\frac{D-n}{2}\right)}  \frac{1}{r^D} \times \nn \\& \times
\left[\frac{A_D}{D-n}\,\left(D\,\frac{ \tx_{M} \tx_{N}}{r^2}-\te_{MN}\right) - (n-1)(D^2-2D-2)
\, \eta_{MN}-2\, \te_{MN} \right].
\end{align}

\subsection{Computation of $\boldsymbol{\xi}$-terms}
Starting with the effective Fourier transforms (\ref{xi0}) and (\ref{dxi}), for the  ${}^{1}T_{MN}$-contributions
 we have explicitly:
\begin{align}\label{qq1}
&T_{MN}^{(0)}\left[\delta{\cal{L}}^{(\xi)}\right]=\bs\int \frac{\dDq }{(2\pi)^{D}}\,\e^{iqx} \delta {\cal{L}}^{(\xi)}(q) \int
  \frac{\dDp }{(2\pi)^{D}}  \frac{ 1}{\left[p^{\kn2}-i\varepsilon\right]\left[(p+q)^2-i\varepsilon\right]} \Bigl(p_M p_N+ q_M  p_N -\frac{1}{2}\,\eta_{MN} \mathbf{q} \mathbf{p}  \Bigr)  \\
&T_{MN}^{(\xi)}\left[\delta{\cal{L}}^{(0)}\right] =2\bs  \int \frac{\dDq \, \dDp}{(2\pi)^{2D}}\,\e^{iqx}  \frac{\delta {\cal{L}}^{(0)}(q,ip) }{\left[p^{\kn2}-i\varepsilon\right]\left[(p+q)^2-i\varepsilon\right]} \bigl(-q_M  p_N +  \eta_{MN} \mathbf{q} \mathbf{p}  \bigr).\label{qq2}
\end{align}
Substituting (\ref{dxi}) into (\ref{qq1}) and integrating over $p$ and $q^{\sigma}$, we obtain:
\begin{align}\label{qq3}
 T_{MN}^{(0)}\left[\delta{\cal{L}}^{(\xi)}\right]= - \bs \frac{ (n-1) (D-2)\, \Gamma(n/2)\,
 \Gamma^2\akn(D/2)\,\Gamma\akn\left(-\frac{D-2}{2}\right)}{  2^{D+1}
      \pi^{(D+n)/2} \Gamma(D)}  \int  d^{\kn n}\akn  \mathbf{q}
       \,\frac{\e^{i\mathbf{qx}}}{ |\mathbf{q}|^{2+n-D}}   \left( \tq_M  \tq_N - |\mathbf{q}|^2 \eta_{MN}    \right).
\end{align}
Substituting (\ref{xi0}) into (\ref{qq2}) and integrating over $p$ and $q^{\sigma}$, one concludes
\begin{align}\label{qq4}
 T_{MN}^{(\xi)}\left[\delta{\cal{L}}^{(0)}\right] =T_{MN}^{(0)}\left[\delta{\cal{L}}^{(\xi)}\right].
\end{align}
Thus combining (\ref{qq3})  with (\ref{qq4}) and integrating, for the regularized ${\kn}^{1}T_{MN} $  we arrive at
\begin{align}\label{xi1final}
 \langle {\kn}^{1} T_{MN}\rangle_{\rm reg}  =   \bs \frac{ (n-1)(D-2)\, \Gamma(n/2)\,\Gamma^3\akn(D/2)\,\Gamma\akn\left(-\frac{D-2}{2}\right)}{ \pi^{D/2}\Gamma(D) \Gamma\akn\left(-\frac{D-n}{2}\right)  }    \left[\frac{1}{D-n}\left(\te_{MN}- D\frac{\tx_M \tx_N}{r^2}\right)+  \eta_{MN}  \right] \frac{1}{r^D}\,.
\end{align}

\medskip

\textbf{Computation of $\boldsymbol{\xi^2}$-term.} The term under interest here, is given by
\begin{align}
 \langle T_{MN}^{(\xi)}[\delta {\cal{L}}^{(\xi)}] \rangle=2\bs \int \frac{\dDq \,\dDp}{(2\pi)^{2D}}\,\e^{iqx}  \frac{\delta {\cal{L}}^{(\xi)}(q,ip) }{\left[p^{\kn2}-i\varepsilon\right]\left[(p+q)^2-i\varepsilon\right]} \bigl(q_M q_N+ q_M  p_N + \eta_{MN} \,\mathbf{q} \mathbf{p}  \bigr).
\end{align}
Integrating
and substituting it with (\ref{dxi}) into (\ref{Tmndxi_eff}), we obtain:
\begin{align}
 \langle T_{MN}^{(\xi)}[\delta {\cal{L}}^{(\xi)}] \rangle_{\reg} =
  - \bs \frac{ (n-1)(D-1) \, \Gamma(n/2)\,\Gamma^2\akn(D/2)\,\Gamma\akn\left(-\frac{D-2}{2}\right)}{2^{D-1}\pi^{(D+n)/2}\Gamma(D)}
  \int d^n \mathbf{q} \,\frac{\e^{i\mathbf{qx}}}{ |\mathbf{q}|^{2+n-D}}  \left(|\mathbf{q}|^2\kn\eta_{MN}  -\tq_{M}\tq_{N}\right).
\end{align}
Comparing it with  (\ref{qq3}) and taking into account (\ref{qq4}), one concludes:
\begin{align}
  \langle {\kn}^{2} T_{MN}\rangle_{\rm reg}  = -  \frac{2\kn (D-1)}{D-2} \,  \langle {\kn}^{1} T_{MN}\rangle_{\rm reg}  = - \frac{1}{2\kn\xi_D}\,  \langle {\kn}^{1} T_{MN}\rangle_{\rm reg} \,.
\end{align}
so their ratio does not depend on the conical subdimensionality $n$.

Integrating the last Fourier integral, we arrive at
\begin{align}
 \langle {\kn}^{2} T_{MN}\rangle_{\rm reg} = - \bs \frac{2\kn(n-1)(D-1)\kn \Gamma(n/2)\,\Gamma^3\akn(D/2)\,
 \Gamma\akn\left(-\frac{D-2}{2}\right)}{ \pi^{D/2}\Gamma(D)\, \Gamma\akn\left(-\frac{D-n}{2}\right)  }
   \left[\frac{1}{D-n}\left(\te_{MN}- D\frac{\tx_M \tx_N}{r^2}\right)+ \eta_{MN}  \right] \frac{1}{r^D}\,, \nn
\end{align}
therefore the combined regularized contribution of the $\xi-$terms equals
\begin{align}\label{xi1finalss}
 \langle   T_{MN} - {\kn}^{0}T_{MN} \rangle_{\rm reg}  =   \bs \frac{(D-2)\, \Gamma(n/2)\,\Gamma^3\akn(D/2)\,
 \Gamma\akn\left(-\frac{D-2}{2}\right)}{ (n-1)^{-1} \pi^{D/2}\Gamma(D)\, \Gamma\akn\left(-\frac{D-n}{2}\right)\,r^D  }
    \left[\frac{1}{D-n}\left(\te_{MN}- D\frac{\tx_M \tx_N}{r^2}\right)+  \eta_{MN}  \right] \xi  \left(1-\frac{\xi}{2 \kn \xi_D} \right).
\end{align}

\subsection{Summary}

Combining (\ref{Tmn_ren}) and (\ref{xi1finalss}), we obtain for the regularized value of energy-momentum VEV:
\begin{align}\label{xicomb1}
 \langle T_{MN} \rangle_{\reg}= \frac{C \mu^{2\varepsilon} \bs}{ r^D}
& \biggl[ \left( \frac{8\kn (D-1)(n-1)}{D-n}\,(\xi-\xi_D)^2+\frac{1}{D^2-1} \right)\!\!
\left(D\,\frac{ \tx_{M} \tx_{N}}{r^2}-\te_{MN}-(D-n)\kn \eta_{MN}\right)+ \frac{\eta_{MN}-\te_{MN}}{D+1}
 \biggr]
\end{align}
with
$$C=\frac{ \Gamma(n/2)\,\Gamma^3\akn(D/2)\,\Gamma\akn\left(-\frac{D-2}{2}\right)}{ 4\kn \pi^{D/2} \,\Gamma(D)\,\Gamma\akn\left(-\frac{D-n}{2}\right)}\,.$$

We see that the classification on parity is based on the factor
$\Gamma\akn\left(-\frac{D-2}{2}\right)\akn/\Gamma\akn\left(-\frac{D-n}{2}\right)$.
Given that \mbox{$d-n\geqslant 1$}, the first pole of $\Gamma-$function in denominator happens at \mbox{$d=n+2$},
 we return exactly to the same dimensionality splitting as for $\langle\phi^{\kn 2}\rangle_{\reg}$.

Hereafter it is more useful to    consider the non-vanishing components of $T_{MN}$ separately:

\begin{enumerate}
  \item The regularized vacuum energy density $\langle T_{00}(x)\rangle_{\reg}$  (as well as flat-sector
  spatial diagonal components $\langle T_{\alpha\alpha}(x)\rangle$):
      \begin{align}\label{Tmn_ren1}
\langle T_{00} (x)\rangle_{\reg}=
\frac{\mu^{2\varepsilon}C (n-1) \bs}{ r^D} \left[8\kn(D-1)\kn (\xi-\xi_D)^2-\frac{1}{D^{\kn2}-1}\right]=-\langle T_{\alpha\alpha} (x)\rangle_{\reg}
\,;\end{align}

  \item The   conical-subspace components $\langle T_{ik}(x)\rangle_{\reg}$
\begin{align}\label{xicomb1mono}
 \langle T_{ik}(x)\rangle_{\reg}= \frac{C \mu^{2\varepsilon} \bs}{ r^{D+2}}
 & \biggl[ \left( \frac{8\kn (D-1)(n-1)}{D-n}\,(\xi-\xi_D)^2+\frac{1}{D^2-1} \right) \Bigl(D\, x_{i} x_{k} -(D-n+1)\,r^2\delta_{ik}\Bigr)
 \biggr].
\end{align}
\end{enumerate}

With respect to the parity of $D$ and $n$ one distinguishes the following cases:

\medskip

\noindent $\bullet$\;\,\textbf{$\boldsymbol{d}$  even,
$\boldsymbol{n}$ odd.} The regularization removal (\ref{Tmn_ren1})  is achieved
in analogy with $\langle\phi^{\kn 2}\rangle_{\reg}$: the pole of the Gamma-function $\Gamma\akn\left(-\frac{D-2}{2}\right)$ in  numerator gives rise to
the corresponding divergent part (as $\epsilon \to 0$)
\begin{align}
 &\langle T_{MN}(x)\rangle_{\rm div}= \frac{(-1)^{d/2+1} \Gamma(n/2)\,\Gamma^2(d/2)}{ 4 \pi^{d/2} \Gamma(d)\,\Gamma\akn\left(-\frac{d-n}{2}\right)}  \frac{  \bs}{ r^d}\frac{1}{\epsilon} \,\Theta_{MN} \\
&\Theta_{MN}  \equiv  \biggl[ \left( \frac{8\kn (d-1)(n-1)}{d-n}\,(\xi-\xi_d)^2+\frac{1}{d^{\kn 2}-1} \right)\! \left(d\,\frac{ \tx_{M} \tx_{N}}{r^2}-\te_{MN}-(d-n)\kn \eta_{MN}\right)   +\frac{1}{d+1} \bigl( \eta_{MN}-\te_{MN}\bigr)
 \biggr],\nn
\end{align}
 and to finite logarithmic and non-logarithmic terms.

In order to reveal the finite part, we have to point out the following observation: as we seen in the Section \ref{phi2ren}, the divergent part corresponding to the pole of a Gamma-function, is accompanied with the logarithmic term in the finite part, and there is some arbitrariness in the non-logarithmic term, related with the finite renormalization of logarithmic scale factor. Here we renormalize the tensor quantity, but the  Gamma-function  $\Gamma\akn\left(-\frac{D-2}{2}\right)$ , which gives a pole, sits in the \textit{common} factor $C$ in (\ref{xicomb1}), while the tensor part is regular. Also taking into account that the finite logarithmic shift due to expansion of $C$ is also common for the whole tensor, we expand $C$ in $\epsilon$ independent of the tensor structure, thus we have the \textit{unified} logarithmic scale factor $\tilde{\mu}$  for \textit{all} components of $ T_{MN}$, while the tensor part in (\ref{xicomb1}) has to be expanded additionally.

Thus for the renormalized tensor we write generically
\begin{align}\label{xicomb1g}
 \langle T_{MN}(x)\rangle_{\ren}= \frac{(-1)^{d/2-1} \Gamma(n/2)\,\Gamma^2\akn(d/2)}{4\kn \pi^{d/2} \,\Gamma(d)\,\Gamma\akn\left(-\frac{d-n}{2}\right)}
   \frac{  \bs}{ r^d}  \Bigl[2\kn \Theta_{MN}  \,\ln \tilde{\mu}r+ A_{MN} \Bigr]\kn.
\end{align}
It also allows the logarithmic-scale finite shift, but within the scalar transformation. In other words, for the scale change  $\mu\to \mu'$ there is an uniparametrical arbitrariness in $A_{MN}$ in the generic form
\begin{align}\label{scalech}
A_{MN}'=  A_{MN} +  2 \kn\Theta_{MN}  \ln \frac{\mu}{\mu'}\,.
\end{align}

Expanding
 \begin{align}
  \frac{8\kn (n-1)\kn(\xi-\xi_D)^2}{(D-1)^{-1}(D-n)}+\frac{1}{D^2-1}  = \left[\frac{8\kn (n-1)\kn(\xi-\xi_d)^2}{(d-1)^{-1}(d-n)}+\frac{1}{d^{\kn 2}-1}\right] +\biggl[\left(\frac{4\kn (n-1)(\xi-\xi_n)}{d-n}\right)^{\!\akn 2}-\frac{1}{(d+1)^2}\biggr]\varepsilon + \mathcal{O}\kn (\varepsilon^2)\nn
  \end{align}
and fixing logarithmic scale as before (as $\tilde{\mu}$, implying the absorbtion of all $D-$dependent coefficients in $C$) one obtains
\begin{align}\label{AMN}
 A_{MN} =  &\left((4\xi-1)^2-\frac{1}{(d+1)^2}  \right)\!(n-1)\kn \eta_{MN} - \biggl[\left(\frac{4\kn (n-1)(\xi-\xi_n)}{d-n}\right)^{\!\akn 2}+\frac{1}{(d+1)^2}\biggr]\te_{MN} + \nn \\
&\;+ \biggl[n \left(\frac{4\kn (n-1)(\xi-\xi_d)}{d-n}\right)^{\!\akn 2}  -(n-1)(4\xi-1)^2+\frac{1}{(d+1)^2} \biggr] \frac{ \tx_{M} \tx_{N}}{r^2}\,.
\end{align}

For the renormalized  vacuum energy density  we obtain:
 \begin{align}
 \langle T_{00} (x) \rangle_{\ren}=\bs \frac{(-1)^{ d/2} (n-1)\,
\Gamma(n/2)\,\Gamma^2(d/2)}{ 4\kn \pi^{ d/2}
\,\Gamma(d)\,\Gamma\akn\left(-\frac{d-n}{2}\right) }
\left[  \left(\frac{2}{d^{\kn 2}-1}-16\kn(d-1)\kn (\xi-\xi_d)^2\right)\ln \tilde{\mu}
r +(4\xi-1)^2-\frac{1}{(d+1)^2}\right]\frac{1}{r^{d}}\kn .\nn
\end{align}

Not hard to conclude that for the values of a curvature-coupling
$$\xi=\xi_d \pm \frac{1}{d-1}\sqrt{\frac{1}{8\kn(d+1)}}$$
the renormalized density $\langle T_{00} (x) \rangle_{\ren}$ does not contain the logarithmic term and thereby does not depend upon the arbitrary constant $\tilde{\mu}$, while the divergent part vanishes: $\langle T_{00} (x) \rangle_{\rm div}=0\, .$

The renormalized  $\langle T_{ik}(x)\rangle$ reads:
\begin{align}\label{xicomb1monoren}
 \langle T_{ik}(x)\rangle_{\ren}=  \bs  \frac{ (-1)^{d/2-1} \Gamma(n/2)\,\Gamma^2(d/2) }{ 2\kn \pi^{d/2} \,\Gamma(d)\,\Gamma\akn\left(-\frac{d-n}{2}\right)}
 & \biggl[ \left( \frac{8(d-1)(n-1)}{d-n}\,(\xi-\xi_d)^2+\frac{1}{d^2-1} \right)\times \nn \\ & \times \Bigl(d\, x_{i} x_{k} -(d-n+1)\,r^2\delta_{ik}\Bigr)\ln \tilde{\mu} r +\frac{1}{2}\,A_{ik}
 \biggr]\frac{ 1}{ r^{d+2}}\,.
\end{align}

In the case (4,3) the expression (\ref{xicomb1g}) reduces to
\begin{align}\label{xi_43}
 \langle T_{MN}\rangle_{\ren}=  \frac{\bs}{8\kn \pi^{2}  r^4}
 & \Biggl[\biggl[ \left( 8\kn \Bigl(\xi-\frac{1}{6}\Bigr)^2+\frac{1}{90} \right)\! \Bigl(4\,\frac{ \tx_{M} \tx_{N}}{r^2}-\te_{MN}-   \eta_{MN}\Bigr)
  +\frac{\eta_{MN}-\te_{MN}}{30}
 \biggr]\ln \tilde{\mu}r+\frac{1}{12}\,A_{MN} \Biggl].
\end{align}

Furthermore, due to the (theoretical) arbitrariness of the constant $\tilde{\mu}$, the non-logarithmic $\xi^2-$terms may be absorbed by the logarithm, introducing the  new constant $\tilde{\mu}'$:
 \begin{align}
 \langle T_{00} (x) \rangle_{\ren}=\bs \frac{(-1)^{ d/2} (n-1)\,
\Gamma(n/2)\,\Gamma^2(d/2)}{ 2\kn \pi^{ d/2}
\,\Gamma(d)\,\Gamma\akn\left(-\frac{d-n}{2}\right) }
\left[  \left(\frac{1}{d^{\kn 2}-1}- 8\kn(d-1)\kn (\xi-\xi_d)^2\right)\ln \tilde{\mu}'
r  -\frac{4\kn \xi }{d-1} +\frac{d^{\kn 3}-1}{(d^{\kn 2}-1)^2} \right]\frac{1}{r^{d}} .\nn
\end{align}
In accord with (\ref{scalech}), this finite shift $\tilde{\mu}\to \tilde{\mu}'= \e^{-1/(d+1)}\tilde{\mu}$ generates the corresponding shift $A_{ik}\to A'_{ik}$ of the spatial (in the conical sector) components.

 For higher-dimensional monopole  ($n=d-1$) equation (\ref{xicomb1g})
reduces to
 \begin{align}
&\langle T_{00} (x) \rangle_{\ren}=  (-1)^{\kn d/2-1}
 \bs \frac{ (d-2)\, \Gamma(d/2)}{2^{d} \pi^{d/2} \,(d-1)  }
\frac{1}{r^{d}} \left[  \left(\frac{1}{d^{\kn 2}-1}- 8\kn(d-1)\kn (\xi-\xi_d)^2\right)\ln \tilde{\mu}'
r  -\frac{4\kn \xi }{d-1} +\frac{d^{\kn 3}-1}{(d^{\kn 2}-1)^2} \right], \nn \\
& \langle T_{ik}(x)\rangle_{\ren}=  \bs  \frac{ (-1)^{d/2} \Gamma(d/2) }{(4\pi)^{d/2} (d-1)}
  \biggl[ \left( 8(d-1)(d-2) \,(\xi-\xi_d)^2+\frac{1}{d^{\kn 2}-1} \right)\! \Bigl(d\, x_{i} x_{k} -
  2\kn r^2\delta_{ik}\Bigr)\ln \tilde{\mu}' r +\frac{1}{2}\,A'_{ik}
 \biggr]\frac{ 1}{ r^{d+2}}\,. \nn
\end{align}
so in the most important  particular case of the spacetime  $(4, 3)-$type it is given by
 \begin{align}\label{Tmn_ren4D}
&\langle T_{00} (x) \rangle_{\ren}=  \frac{\bs }{ 4
\pi^{2} } \left[  \left(  4\kn \Bigl(\xi- \frac{1}{6}\Bigr)^2
-\frac{1}{90}\right)\ln \mu'r +\frac{2}{9}\Bigl(\xi- \frac{21}{100}\Bigr) \right]\frac{1}{r^4}\,,\nn \\
 & \langle T_{ik}(x)\rangle_{\ren}=   \frac{\bs}{4\pi^{2}}
\biggl[ \left(8\Bigl(\xi- \frac{1}{6}\Bigr)^{\!2}+\frac{1}{90} \right)\! \Bigl(2\, x_{i} x_{k} -  r^2\delta_{ik}\Bigr)\ln \tilde{\mu}' r +\frac{1}{24}\,A'_{ik}
 \biggr]\frac{ 1}{ r^{6}}\,.
\end{align}

Now we can  compare our result (\ref{xi_43}) with the linear-in-$\bs$ part of the corresponding expression in \cite{Mazzitelli}, applied  to the spacetime-at-hand.

The logarithmic expression in \cite{Mazzitelli} within our accuracy\kn\footnote{It implies that we have neglected $\mathcal{O}\kn(R^2)-$terms.}
generically is given by
\begin{align}\label{Mazzone}
\langle T_{MN}(x)\rangle_{\rm log}=\frac{1}{{160}\pi^2}\left[\left(\frac{1}{3}-\frac{10}{3}\,\xi+10\xi^2  \right) R_{\cd \akn MN}-\frac{1}{6}\,
\Box R_{MN} + \left(-\frac{1}{4}+\frac{10}{3}\,\xi-10\xi^2\right)g_{MN}\,\Box R \right]\ln \mu r,
\end{align}
while the non-logarithmic one is arbitrary.
Substituting the Ricci tensor and Ricci-scalar (\ref{Riccies}),
and making use of
$$R_{\kn\cd \akn MN}=\frac{4\kn(1-\beta^2)}{r^4}\Bigl(4 \frac{\tx_{M}\tx_{N}}{r^2}-\te_{MN}  \Bigr), \qquad\qquad \Box R_{ MN}= 4 \kn(1-\beta^2)\frac{\tx_{M}\tx_{N}}{r^6}\,, $$
 one concludes that our  expression (\ref{xi_43}) has a discrepancy with (\ref{Mazzone}) by factor of two, for all monomials $\eta_{MN}$, $\te_{MN}$  and $\tx_{M}\tx_{N}$, respectively. Meanwhile, the corresponding expression for $\langle\phi^{\kn 2}\rangle$ perfectly matches.
Such a discrepancy implies necessity of re-derivation of the generic expression in the work \cite{Castagnino} (actually referred by \cite{Mazzitelli}). Following their ideology, based on the de\kn{}Witt-Schwinger kernel, we could fix some inaccuracy of these works\footnote{Actually, the pre-logarithmic  expression (\ref{Mazzone}), multiplied by 2, coincides with the  pre-logarithmic coefficient in logarithmically-divergent part of the corresponding expression by Christensen \cite{christ76} for renormalized VEV for \textit{massive} scalar field's energy-momentum tensor.}. Thus we think that if take into account the fixing coefficient, our result (\ref{xi_43}) coincides with the generic one in the logarithmic term, whereas it contains information about the non-logarithmic term.

\medskip

\noindent $\bullet$\;\,\textbf{$\boldsymbol{d}$ and
$\boldsymbol{n}$ odd.} Now  $\Gamma\akn\left(-\frac{D-2}{2}\right)$ in the numerator is regular, while $\Gamma\akn\left(-\frac{D-n}{2}\right)$ in the denominator is infinite, hence the total  renormalized  $\langle T_{MN}(x)\rangle $ vanishes:
\begin{align}
\langle T_{MN}(x)\rangle_{\ren}= 0\,,
\end{align}
in accord with the corresponding value of $\langle\phi^{\kn 2}\rangle$ \footnote{See the footnote \ref{footinootie} on page \pageref{footinootie}.}.

\medskip

\noindent $\bullet$\;\,\textbf{$\boldsymbol{d}$  odd,
$\boldsymbol{n}$ even.}  Here both  $\Gamma\akn\left(-\frac{D-2}{2}\right)$ in the numerator and
 $\Gamma\akn\left(-\frac{D-n}{2}\right)$ in the denominator are regular,
 with semi-integer arguments, so $\langle \kn T_{MN}(x) \rangle_{\rm div}=0$ and we have simply
\begin{align}
\langle T_{MN}(x)\rangle_{\ren}= \frac{ \Gamma(n/2)\,\Gamma^3\akn(d/2)\,\Gamma\akn\left(-\frac{d-2}{2}\right)}{ 4\kn \pi^{d/2} \,\Gamma(d)\,\Gamma\akn\left(-\frac{d-n}{2}\right)}
\frac{  \bs}{ r^d} \, \Theta_{MN}\,. \nn
\end{align}

Transforming it with help of (\ref{gammae}), one obtains
 \begin{align}\label{p1}
\langle T_{MN}(x)\rangle_{\ren}= (-1)^{n/2-1}\frac{ \Gamma(n/2)\,\Gamma^2\akn(d/2)\,\Gamma\akn\left(\frac{d-n+2}{2}\right)}{ 4\kn \pi^{d/2} \,\Gamma(d)}
\frac{  \bs}{ r^d} \, \Theta_{MN}\,.
\end{align}

In particular, for the $d-$dimensional monopole ($n=d-1$) the renormalized energy-momentum tensor reads:
 \begin{align}
\langle T_{MN} \rangle_{\ren}= \frac{\pi^{1-d/2} \kn \Gamma(d/2)  }{ (-4)^{(d+1)/2} }
\frac{  \bs}{ r^d} \biggl[ \!\left(  8 (d-2) \,(\xi-\xi_d)^2+\frac{d+1}{(d^{\kn 2}-1)^2} \right)\! \!\left(d\,\frac{ \tx_{M} \tx_{N}}{r^2}-\te_{MN}- \eta_{MN}\right)   +\frac{ \eta_{MN}-\te_{MN}}{d^{\kn 2}-1}
 \biggr].\nn
\end{align}

\medskip

\noindent $\bullet$\;\,\textbf{$\boldsymbol{d}$ and
$\boldsymbol{n}$ even.} Here both  $\Gamma\akn\left(-\frac{D-2}{2}\right)$ in the numerator and  $\Gamma\akn\left(-\frac{D-n}{2}\right)$ in the denominator are singular, so their ratio is determined by the ratio of corresponding residuals (\ref{resi}).

Thus $\langle T_{MN}(x) \rangle_{\rm div}=0$, and
 \begin{align}\label{p2}
\langle T_{MN}(x)\rangle_{\ren}= (-1)^{n/2-1}\frac{ \Gamma(n/2)\,\Gamma^2\akn(d/2)\,\Gamma\!\left(\frac{d-n+2}{2}\right)}{ 4\kn \pi^{d/2} \,\Gamma(d)}
\frac{  \bs}{ r^d} \, \Theta_{MN}\,.
\end{align}

Again, the formulae (\ref{p1}) and (\ref{p2}) are identical, and represent the unified expression for even $n$, like it was for $\langle\phi^{\kn 2}\rangle$.

\medskip

Summarizing, in this section we have computed the renormalized
vacuum averaged  $\langle T_{MN}\rangle_{\ren}$ of the massless scalar field in the background of  (global) monopole up to the first order in $\bs$.
Computing along the same ideology as in previous section, we obtain the same splitting with respect to the parity of a dimensionalities of the total spacetime  and its deficit-angle submanifold.
Here  the most actual case with even $d$ and odd $n$ (in particular, for the (4,3)-type of a spacetime) demands the more accuracy working with logarithms, due to the tensorial structure of  $\langle T_{MN}\rangle_{\reg}$. The logarithmic mass-scale change generates the uniparametric equivalence class of the non-logarithmic symmetric  tensors  $A_{MN}$\footnote{Contrary to the result of \cite{Mazzitelli} where this matrix is symmetric but arbitrary.}, representing the linear shell of monomials $\eta_{MN}$, $\te_{MN}$ and $\tx_{M}\tx_{N}$. For definite value of $\xi$, the logarithmic term and corresponding logarithmic uncertainty can be removed from $\langle T_{00}\rangle_{\ren}$. However, contrary to the case of $T_{00}$, no value of coupling $\xi$ kills the logarithmic term arising in  $\langle T_{ik}\rangle_{\ren}$
since both terms in the parenthesis of (\ref{xicomb1mono}) are positive.
Finally, no value of $\xi$ eliminates the logarithmic arbitrariness both in $\langle\phi^{\kn 2}\rangle_{\ren}$ and in $\langle T_{MN}\rangle_{\ren}$ simultaneously.

The other cases of $d$ and $n$ are similar to those ones of $\langle\phi^{\kn 2}\rangle_{\ren}$.

In the next section we show that the Green's function obtained above,
 enables  to consider the well-known purely classical problem of a
gravity-induced self-action on a charge placed at fixed point of
the space under consideration.

\section{Static self-energy and self-force of a pointlike charge}\label{Self-force}

As it was concluded in (\ref{e6qq}) and (\ref{emse}), the
self-energy of a scalar $(q)$ or electric $(e)$ point charge in an
ultrastatic $d-$dimensional spacetime is determined by the
coincidence-limit of the Euclidean Green's function on the spacetime
with the dimensionality $(d-1)$:
\begin{equation}
U_{\sc}(x)=\frac{q^2}{2}\, \kn G^E_{\ren}(x, x\mi d-1, n)\,,
\qquad\qquad   U_{\em}(x)= \frac{e^2 }{2}\,\kn G_{\ren}^E(x, x\mi
d-1, n)\Bigl|_{\kn\xi=0}\,.
\end{equation}
The relation between self-energy and the self-force is given by (\ref{e6}). Taking into account
that for the self-energy  the first non-vanishing order is $\mathcal{O}\kn(\bs)$, one obtains
to the lowest order simply
$$\mathbf{F}_{\ren}=-{\rm grad} \,U_{\ren}\, .$$
Moreover, simple relation between scalar and electrostatic
self-energy (\ref{emse}) enables  to restrict the consideration
by the scalar one.

According to (\ref{g4}), the regularized scalar gravity-induced
self-energy is given by
\begin{align}\label{g4ss}
U_{\reg}=  q^2
 \mu^{2\varepsilon} \bs
\frac{n-1}{8\kn \pi^{(D-1)/2}}
\frac{\Gamma(n/2)\,\Gamma^3\!\left( \frac{D-1}{2}\right)}{\Gamma(D-1)}\Bigl(\frac{\xi}{\xi_{D-1}}-1\Bigr)\,\frac{\Gamma\!\left(-\frac{D-3}{2}\right)}{\Gamma
\bigl(\frac{3-(D-n)}{2}\bigr)}\,\frac{1}{r^{D-3}}\,.
\end{align}

Now the classification is determined basically by the factor
\begin{align}\label{tt0}
\Gamma\akn\Bigl(-\frac{D-3}{2}\Bigr)\akn\Bigl/\Gamma
\akn\Bigl(\frac{3-D+n}{2}\Bigr)\kn .
\end{align}

With respect to the parity of $d$ and $n$ one distinguishes the
following cases:

\medskip

\noindent $\bullet$\;\,\textbf{$\boldsymbol{d}$  even,
$\boldsymbol{n}$ odd.} The Gamma-function  $\Gamma\akn\bigl(-\frac{D-3}{2}\bigr)$ is regular, while
$\Gamma
\akn\bigl(\frac{3-D+n}{2}\bigr)$ tends to its pole (unless $D-n=1$). Therefore the renormalized self-energy and the self-force vanish generically in this case:
$$U_{\ren}=0\,,  \qquad\qquad {\mathbf{F}}_{\ren}=0\,.  $$

For the  exceptional case $d-n=1$ both Gamma-functions are regular, hence
\begin{align}\label{tt1}
 U_{\ren}= -q^2 \bs \,(-1)^{d/2}\,
\frac{d-2 }{8\kn \pi^{(d-3)/2}} \frac{ \Gamma^3\!\left(
\frac{d-1}{2}\right)}{\Gamma\left( d-1\right)}
\Bigl(\frac{\xi}{\xi_{d-1}}-1\Bigr)\,\frac{1}{r^{d-3}}\,.
\end{align}
The corresponding self-force is given by
\begin{align}\label{tt2}
 \mathbf{F}_{\ren}=- q^2 \bs \,(-1)^{d/2}\,
\frac{(d-2)(d-3)}{8\kn \pi^{(d-3)/2}} \frac{ \Gamma^3\!\left(
\frac{d-1}{2}\right)}{\Gamma\left(
d-1\right)}\Bigl(\frac{\xi}{\xi_{d-1}}-1\Bigr)\,\frac{
\mathbf{r}}{r^{d-1}}\,.
\end{align}
Thus, at $\xi=\xi_{d-1}=(d-3)/4(d-2)$ the renormalized self-energy
and self-force vanish.

In particular case of the $(4,3)$-spacetime one obtains
\begin{align}\label{tt3}
 U_{\ren}=-q^2 \bs \,
\frac{\pi \kn (8\kn {\xi} -1  ) }{2^{6} r}\,, \qquad\qquad
\mathbf{F}_{\ren}=-q^2 \bs \, \frac{\pi \kn ( {8\kn \xi} -1 )
}{2^{6} } \,\frac{ \mathbf{r}}{r^{3}}\,.
\end{align}
 Thus, the  pointlike charge feels the monopole as a point charge with the magnitude $2^{-4}\beta'(8
\kn {\xi} -1)\,\pi^2 q$ localized at the point $r=0$. For values
$\xi>\xi_3=1/8$ the self-force is attractive (in particular, for
the conformal coupling, $\xi=\xi_4=1/6$), while for values
$\xi<1/8$ the self-force is repulsive.

In the case of electrostatic self-action (according to the eq.\,(\ref{emse}) one has to put $\xi=0$ in (\ref{tt3}) and replace
$q^2$ by $e^2$) our result (\ref{tt3}) coincides with the one of
the paper \cite{Bezerra_97}.

\medskip

\noindent $\bullet$\;\,\textbf{$\boldsymbol{d}$ and
$\boldsymbol{n}$ odd.} In this case the Gamma-function
$\Gamma\akn\bigl(-\frac{D-3}{2}\bigr)$ is singular, while
$\Gamma\akn\bigl(\frac{3-D+n}{2}\bigr)$ is regular. This leads to
the non-zero diverging part, and the finite renormalized value of
the self-energy takes the form
\begin{align}\label{tt4}
U_{\ren}=(-1)^{(n+3)/2} q^2 \bs \frac{n-1}{8\kn \pi^{(d+1)/2}}
\frac{\Gamma(n/2)\,\Gamma^2\!\left( \frac{d-1}{2}\right) \,\Gamma
\bigl(\frac{d-n-1}{2}\bigr)}{\Gamma(d-1) }
\left[\Bigl(\frac{\xi}{\xi_{d-1}}-1\Bigr)\ln \tilde{\mu}r +
\frac{1}{(d-2)(d-3)}\right]  \frac{1}{r^{d-3}}
\end{align}
with arbitrary $\tilde{\mu}$.

The corresponding self-force reads
\begin{align}\label{tt5}
\mathbf{F}_{\ren}=(-1)^{(n+3)/2} q^2\bs \frac{n-1}{8\kn
\pi^{(d+1)/2}} \frac{\Gamma(n/2)\,\Gamma^2\!\left(
\frac{d-1}{2}\right) \,\Gamma
\bigl(\frac{d-n-1}{2}\bigr)}{\Gamma(d-1)
}\left[\Bigl(\frac{\xi}{\xi_{d-1}}-1\Bigr)\bigl((d-3)\ln
\tilde{\mu}r -1\bigr)+ \frac{1}{(d-2)} \right]  \frac{
\mathbf{r}}{r^{d-1}}\,.
\end{align}
 For $\xi=\xi_{d-1}$ the result becomes free of uncertainty.

\medskip

\noindent $\bullet$\;\,\textbf{$\boldsymbol{d}$  odd,
$\boldsymbol{n}$ even.} Here the Gamma-function
$\Gamma\akn\bigl(-\frac{D-3}{2}\bigr)$ is singular, while
$\Gamma\akn\bigl(\frac{3-D+n}{2}\bigr)$ is also singular, unless
$d=n+1$. Hence, in the generic case the divergent part of the
self-energy vanishes, and $U_{\ren}$  is determined by the ratio
of corresponding residuals:
\begin{align}\label{tt6}
U_{\ren}=  (-1)^{n/2}q^2 \bs \frac{n-1}{8\kn \pi^{(d-1)/2}}
\frac{\Gamma(n/2)\,\Gamma^2\!\left( \frac{d-1}{2}\right)  \Gamma
\bigl(\frac{d-n-1}{2}\bigr)}{\Gamma(d-1)}\Bigl(\frac{\xi}{\xi_{d-1}}-1\Bigr)\,
\,\frac{1}{r^{d-3}}\, .
\end{align}
Corresponding self-force equals
\begin{align}\label{tt7}
\mathbf{F}_{\ren}=  (-1)^{n/2} q^2 \bs \frac{(n-1)(d-3)}{8\kn
\pi^{(d-1)/2}} \frac{\Gamma(n/2)\,\Gamma^2\!\left(
\frac{d-1}{2}\right)  \Gamma
\bigl(\frac{d-n-1}{2}\bigr)}{\Gamma(d-1)}\Bigl(\frac{\xi}{\xi_{d-1}}-1\Bigr)
\,\frac{\mathbf{r}}{r^{d-1}}\,.
\end{align}
In the exceptional case of the higher-dimensional monopole
($d=n+1$) the denominator $\Gamma\akn\bigl(\frac{3-D+n}{2}\bigr)$
is regular, hence we return to the logarithmic case: along the
same lines  as previously we obtain
\begin{align}\label{tt8}
& U_{\ren}=q^2
 \bs
\frac{(-1)^{(d+1)/2}}{8\kn \pi^{(d-1)/2}}
\frac{ \Gamma^2\!\left( \frac{d-1}{2}\right)}{\Gamma(d-1)}\left[(d-2)\Bigl(\frac{\xi}{\xi_{d-1}}
-1\Bigr)\ln \tilde{\mu}r+  \frac{1}{d-3}\right] \frac{1}{r^{d-3}}\, , \\
& \mathbf{F}_{\ren}=q^2
 \bs
\frac{(-1)^{(d+1)/2}}{8\kn \pi^{(d-1)/2}} \frac{ \Gamma^2\!\left(
\frac{d-1}{2}\right)}{\Gamma(d-1)}\left[(d-2)\Bigl(\frac{\xi}{\xi_{d-1}}-1\Bigr)\bigl((d-3)\ln
\tilde{\mu}r-1\bigr)+  1\right] \frac{\mathbf{r}}{r^{d-1}}\,.
\end{align}

\medskip

\noindent $\bullet$\;\,\textbf{$\boldsymbol{d}$ and
$\boldsymbol{n}$ even.} Here both Gamma-functions in  (\ref{tt0})
are regular, hence the  divergent part vanishes and after
transformations with the help of (\ref{gammae}) we have just
\begin{align}\label{tt9}
&U_{\ren}= (-1)^{n/2}q^2 \bs \frac{n-1}{8\kn \pi^{(d-1)/2}}
\frac{\Gamma(n/2)\,\Gamma^2\!\left( \frac{d-1}{2}\right)\,\Gamma\!\left(\frac{d-n-1}{2}\right)}
{\Gamma(d-1)}\Bigl(\frac{\xi}{\xi_{d-1}}-1\Bigr)\,\frac{1}{r^{d-3}}\, , \nn \\
& \mathbf{F}_{\ren}=  (-1)^{n/2} q^2 \bs \frac{(n-1)(d-3)}{8\kn
\pi^{(d-1)/2}} \frac{\Gamma(n/2)\,\Gamma^2\!\left(
\frac{d-1}{2}\right)  \Gamma
\bigl(\frac{d-n-1}{2}\bigr)}{\Gamma(d-1)}\Bigl(\frac{\xi}{\xi_{d-1}}-1\Bigr)
\,\frac{\mathbf{r}}{r^{d-1}}\,.
\end{align}

\medskip

\textit{To summarize}: based on the formal relation of the Feynman propagator
with Euclidean Green's function in the
 coincidence-point limit, we have expressed the regularized self-action via regularized Green's function of
 the previous dimensionality.  As before, the consideration splits onto four characteristic cases
 of parities $d$ and $n$,
 though here one meets the significant exceptions of the monopole background with no flat spatial dimensions ($n=d-1$).
In the most cases the self-action looks like the flat-space Coulomb
interaction of a charge $q$ with a charge \mbox{$\sim  (\xi-\xi_{d-1})\,q$}  placed into the monopole position,
 and vanishes for the particular value $\xi=\xi_{d-1}$ of the curvature coupling.
In the case of odd $d$ while $n$ is odd or equal to $d-1$, there is
an additional logarithmic multiplier, which depends on the
arbitrary parameter $\tilde{\mu}$.

Finally, comparing (\ref{tt9}) with (\ref{tt6}) and (\ref{tt7}),  we notice that for even $n$
the cases with even and odd $d$
 can be combined into the unified formula (except for the full-hyperspace monopole case), in accord with the
 previous computations of the
 renormalized $\langle\phi^{\kn 2}\rangle$ and $\langle T_{MN}\rangle$.

\section{Vacuum polarization near cosmic string revisited}\label{Cosmostring}

Now consider the particular case of a two-dimensional $(n=2)$
conical subspace. If $d=3\kn(4)$ this space is  the spacetime of a
point mass (infinitely thin straight cosmic string).

This problem was considered in a series of papers. The primary
goal of our consideration is to show that there is some ambiguity
in previous calculations in the case of a non-minimally coupled
massless scalar field.

Indeed, in  calculations
\cite{Helliwell,Linet_1986,Frolov_1987,Dowker,Sakhni_1992,Grats1995}
the starting point is the expression (\ref{putinmerde}) with
operator ${D}_{MN}$, whose form is determined by the classical
expression for the energy-momentum tensor and thus includes the
$\xi-$dependent terms. Whereas as a Green's function the authors used
the Green's function for a \textit{minimally coupled} scalar field.
Thus, it was supposed that one can extract a $\delta^2-$like
potential from the wave equation, arguing it by the fact that the
space is flat everywhere outside the point mass/cosmic string.
This Green's function does not depend on $\xi$ and in the limit
$\beta\rightarrow 1$ tends to the flat Green's function $G^F_0(x-x')$,
which is the solution of the equation
\begin{align}
\eta^{MN}\partial_M\partial_N\,G_0^F(x-x')=-\delta^d(x-x')\, .
\end{align}

On the other hand, we can start from the explicit equation
\begin{align}\label{gfold}\sqrt{-g}\,\bigl[\Box-\xi R\bigr]G^F_{\xi}(x,
x')=-\delta^d(x-x')\, .\
\end{align}
 In the coordinates of usage here, the potential reads
\begin{align} \label{pot2}
\gamma(r)=\sqrt{-g}\,\xi R=4\pi\bs\xi\,\delta^2(\mathbf{r})\,
,\quad \mathbf{r}=(x^1, x^2)\, .
\end{align}
 In the Eq. (\ref{gfold}) there
are two independent parameters, namely $\bs$ and $\xi$. Suppose,
that there exists a limit of the Green's function $G^F_{\xi}$,
when $$\bs\rightarrow 0\, ,\qquad \xi\rightarrow\infty\, ,\qquad
\lambda \equiv   4\pi\,\xi\,\bs={\rm const}\, .$$ Let us denote it as
$G_{\lambda}^F$. In this limit Eq. (\ref{gfold}) takes the form
\begin{align}\label{gfold2}\left[\eta^{MN}\partial_M\partial_N -
\lambda\,\delta^2(\mathbf{x})\right]G^F_{\lambda}(x,
x')=-\delta^d(x-x')\,
 .
 \end{align}
It is obvious that, if the limit does exist, $G_{\lambda}^F$ can
not be equal to the flat-space Green's function $G_0^F$.

The corresponding equation for the scalar field $\phi$ can be reduced
to a stationary two-dimensional Schr\"{o}dinger-like equation with
a planar $\delta^2-$function potential. Equations of this kind
have been widely discussed in the literature. It was shown that
 these interactions
require regularization and infinite renormalization of the
coupling constant and lead to non-trivial physical results. Alternatively, one
can follow more satisfactory approach based on a self-adjoint
extension of a noninteracting Hamiltonian, defined on a space with
one extracted point (see \cite{Albeverio,Jackiw_1995} and Refs
therein).

We think, that the example above demonstrates the necessity to
revise  the vacuum polarization effects on manifolds with
$\delta^2-$like singularities.  This problem demands
consideration in more detail. Here we restrict ourselves by the
consideration of this problem in the framework of the perturbation
approach.

Thus, we start from the expression (\ref{Obama}) with the
potential $\gamma$ defined by the Eq. (\ref{pot2}). The Fourier
transform of this potential  has the form \be\label{fourgamma}
\mathcal{F}[\gamma(r)]=
4\pi\bs(2\pi)^{d-2}\,\delta(q^0)\prod\limits_{N=3}^{d-1}\delta(q^N)
\, . \ee Substituting (\ref{fourgamma}) into Eg.(\ref{Obama}), we
obtain that with our accuracy
\begin{align}\label{g_str} G^F(x, x'\mi d, 2)=G^F_0(x-x')+\frac{\bs}{\pi} \int d^2 q\,
\frac{\e^{i\mathbf{q}\mathbf{x}}}{{\mathbf{q}}^2}\,\int
\frac{\ddp}{(2\pi)^d}\,\e^{ip(x-x')}\,
\frac{{\mathbf{p}}^2-\xi\,{\mathbf{q}}^2}{\left[p^{\kn
2}-i\varepsilon\right]\left[(p+q)^2-i\varepsilon\right]}\,
.\end{align}

Starting from (\ref{g_str}) and proceeding along the same line as
in the previous sections, we obtain:
\begin{align}\label{qq1a}
 \langle\phi^2(x)\rangle_{\ren}=-i\,G^F_{\ren}(x, x\mi d, 2)=
-\frac{\bs}{2 \pi^{d/2}}\biggl(\frac{\xi}{\xi_d}-1\biggr)
\frac{\Gamma^3(d/2)}{(d-2)\kn\Gamma(d)}\,\frac{1}{r^{d-2}}
\end{align}
for the renormalized vacuum expectation value of the field square
and
\begin{align}\label{qq2a}
&\langle T_{00}\rangle_{\ren}=   \bs \frac{ \Gamma^3(d/2)}{4\kn \pi^{d/2} \,\Gamma(d) }
\left(  8\kn (d-1)\kn (\xi-\xi_d)^2
-\frac{1}{d^{\kn2}-1}\right)\frac{1}{r^{d}}\, ,  \nn \\
& \langle T_{11}\rangle_{\ren}=   \bs \frac{ \Gamma^3\akn(d/2) }{
4\kn \pi^{d/2} \,\Gamma(d)} \,\biggl[  \frac{2\kn
(\xi-\xi_d)^2}{\xi_d}+\frac{1}{d^{\kn 2}-1} \biggr] \! \Bigl(
x_1^2
- (d-1)\, x_2^2 \Bigr)\frac{1}{ r^{\kn d+2}}\, , \nn \\
&\langle T_{22}\rangle_{\ren}= \langle
T_{11}\rangle_{\ren}\bigl|_{x_1\leftrightarrow x_2}\, ,\\
& \langle T_{12}\rangle_{\ren}=\langle T_{21}\rangle_{\ren}=\bs
\frac{d\kn\Gamma^3\akn(d/2) }{ 4\kn \pi^{d/2} \,\Gamma(d)}
\,\biggl[  \frac{2\kn (\xi-\xi_d)^2}{\xi_d}+\frac{1}{d^{\kn 2}-1}
\biggr] \frac{x_1 x_2}{ r^{\kn d+2}}\, ,\nn
\\&\langle
T_{\alpha\beta}\rangle_{\ren}=-\delta_{\alpha\beta}\kn\langle
T_{00}\rangle_{\ren}\, ,\quad \alpha, \beta, \pp= n+1, \pp,
d-1\, \nn
\end{align}
for the nonzero components of $\langle T_{MN}\rangle_{\ren}$.

The corresponding classical gravity-induced scalar self-energy and self-force are given by
\begin{align}\label{tt9s}
U_{\ren}=-
q^2 \bs
\frac{ \Gamma^2\!\left( \frac{d-1}{2}\right)\,\Gamma\!\left(\frac{d-3}{2}\right)}{8\kn \pi^{(d-1)/2}\Gamma(d-1)}
\Bigl(\frac{\xi}{\xi_{d-1}}-1\Bigr)\,\frac{1}{r^{d-3}}\,,  \qquad\qquad
 \mathbf{F}_{\ren}=-
\frac{ q^2\bs }{4\kn \pi^{(d-1)/2}}
\frac{ \Gamma^3\!\left( \frac{d-1}{2}\right)  }{\Gamma(d-1)}\Bigl(\frac{\xi}{\xi_{d-1}}-1\Bigr) \,\frac{\mathbf{r}}{r^{d-1}}\,.
\end{align}
Thus in any dimension the self-force is attractive for
$\xi>\xi_{d-1}$,  repulsive vice versa, and equal to zero if
$\xi=\xi_d-1$.

Our results coincide with the ones of the papers
\cite{Grats96,Grats98,Grats00,Helliwell,Linet_1986,Frolov_1987,Dowker,Sakhni_1992,Grats1995}
in the case of a minimally coupled scalar field in the
three-/four-dimensional spacetime, but differ from those if
$\xi\neq 0$. As it was mentioned above, this distinction is a
consequence of the fact that Green's function satisfies Eq.\,(\ref{gfold}). The latter
 contains an
additional  a two-dimensional
$\delta^2(\mathbf{x})\kn$-potential   which was not taken into
account in the cited papers.

 However, our result for ${\langle T_{MN}\rangle}_{\ren}$ coincides with \cite{Helliwell,Frolov_1987} also for the particular value \mbox{$\xi=\xi_4=1/6$}.
 This <<occasional>> coincidence follows from the fact that the sum
 $\langle \kn T_{MN}^{(0)}\bigl[\delta{\cal{L}}^{(\xi)}\bigr]+T_{MN}^{(\xi)}\bigl[\delta{\cal{L}}^{(\xi)}\bigr]\rangle_{\reg}$, representing the discrepancy,
  is proportional to $\xi\kn(\xi-\xi_D)$. If the divergent part
 vanishes, that is the case for the cosmic string, then the latter equals $\xi\kn(\xi-\xi_d)$ and thus vanishes for the conformal coupling also.

Notice, our results (\ref{qq1a}), (\ref{qq2a}) and (\ref{tt9s})
coincide  with the  \mbox{$n\to 2$} limit of the results obtained
in the Sections \ref{phi2ren} and \ref{EMTVEV}. First of all, as
we mentioned, the monopole's results for even $n$ are combined (for odd and even $d$).
Furthermore, we see that the expressions (\ref{g4_www}),
(\ref{g4_even}), (\ref{p1}), (\ref{p2}),  (\ref{tt9}), (\ref{tt7})
and (\ref{tt9}) are regular at \mbox{$n=2$}, though the initial
expression for Ricci-scalar was singular in this limit
(\ref{curvlin}), representing \textit{the only difference}. Next,
we observe that this difference disappears after the first
Fourier-transform. Indeed, the Fourier-transform of the string's
Ricci-scalar  perfectly coincides with the (regular) \mbox{$n\to
2$} limit of the corresponding  Fourier-transform of the
monopole's Ricci-scalar, readily computed with help of
(\ref{Four1}). Since the  rest of computation is the same for the
cosmic string and global monopole, no wonder that we have obtained
coincidence in the final formulae.

\section{Conclusion}\label{Concl}


On curved
backgrounds being  multidimensional generalizations of the well-known four-dimensional
  cosmic string/global monopole, we  have considered  two, disconnected at the first
glance, problems. Namely, the gravity-induced vacuum polarization of a
massless scalar field and the classical   self-action of a static
 scalar or electric charge.
However, the technique to solve the  problems under consideration  turned out to be similar, since it
 refers to getting the compact workable expression for
the Green's function and its derivatives in the coincidence-point
limit for all \mbox{$d\geqslant 3$} and \mbox{$2 \leqslant n \leqslant    d-1$},
representing our primary particular goal. For this purpose we use
the methods of perturbation theory. Taking into account the actual
smallness of the angle deficit,
we have performed computations in  first
order with respect to the  angular deficit value. Since, in principle,
both the vacuum expectation values and the classical self-energy
are divergent, for regularization and renormalization of these
quantities we adapted the dimensional-regularization method.


The zeroth computational order is determined by the Minkowskian
Green's function and completely consists of the tadpole-like
contributions (\ref{tadp}). In quantum field theory, the
appearance of divergences produced by tadpoles, is explained by
the fact that the perturbation theory is constructed with respect
to a nonphysical vacuum, while their elimination is explained by
the necessity of redefining the vacuum state. In the framework of
self-action, it is of interest to understand why similar
divergences appear in the classical theory too. Following the
prescriptions of the quantum field theory, we assumed all
expressions of the form  (\ref{tadp}) to be equal to zero. The
motivation for this recipe is not associated in any way with the
quantum theory. Actually, it relies on the absence of dimensional
parameters in the corresponding expression and, as a consequence,
on the impossibility to assign some reasonable finite value,
except zero, to such integrals under regularization. Therefore,
this rule is also equally applicable within the classical field
theory.


The desired effects are computed in the first in $\bs$ order. Already starting from the Green's function, for all of our computational tasks we meet
the  characteristic ratio of two Gamma-functions, which splits the consideration of all $(d,n) $-types onto four
 characteristic cases, depending on parities of $d$ and $n$. The poles of Gamma-function may arise in numerator,
  in denominator, or in both. However, in the very end of
computation one can combine all formulae with even $n$ (for arbitrary $d$) into the unified case.


With help of the regularized Green's function we have computed the
renormalized vacuum averaged  $\langle\phi^{\kn 2}\rangle_{\ren}$
and  $\langle T_{MN}\rangle_{\ren}$ for a massless scalar field
coupled with the generalized conical background (\ref{Metnew}) via
an arbitrary coupling $\xi$. The expressions for vacuum averaged $\langle\phi^{\kn 2}\rangle_{\ren}$, corresponding to all
characteristic cases (with our accuracy),
 vanish at
\mbox{$\xi=\xi_d$}. In the case with even $d$ and odd $n$ (in particular, for
the (4,3)-type of a
 spacetime) the VEVs of $\langle\phi^{\kn 2}\rangle_{\ren}$ and  $\langle T_{MN}\rangle_{\ren}$ contain
 logarithmic factor. We are in agreement with
 \cite{Bezerra_2006, Mazzitelli} in the pre-logarithmic coefficient. Concerning
 the non-logarithmic term in $\langle T_{MN}\rangle_{\ren}$, we restrict its arbitrariness by  the single arbitrary parameter,
fixing the more wide freedom in \cite{Mazzitelli}.

For the self-action, in addition to the four basic characteristic
cases of parities $d$ and $n$, there is a significant exception
 of the monopole background  (\mbox{$n=d-1$}).
In the most cases the self-action represents the Coulomb-like field with <<charge>> \mbox{$(\xi-\xi_{d-1})$}
  and vanishes for the particular value $\xi=\xi_{d-1}$ of the curvature coupling.
Also it should be mentioned that (for $\xi=0$) the gravity-induced self-energy and the self-force of the
point-like static $electric$ charge  $e$ can be
obtained from our expressions
 by the formal identification $q^2 \to e^2$, since the spacetime-at-hand is ultrastatic, and the defining
 expressions for spatial scalar and vectorial Green's
 functions coincide.



We'd like to emphasize that within our scheme, the appearance of the mass-dimensionful term inside the logarithm
is related neither with the
arbitrary scale factor $r_0$ coming from the cartesian coordinates (\ref{Metnew}), nor with any length/mass
of the problem-at-hand since the latter is absent\footnote{For the real cosmic string one has its real width,
 but the results for the cosmic string
within  our model do not concern logarithms.}. The logarithmic scale factor follows from the regularization
 (\ref{regul}) and its value, in principle,
is arbitrary.

 Making use of the same approach, but applied to the delta-like interaction in the infinitely
 thin straight cosmic string,
 we have computed the effects under consideration. The results coincide with the known in literature
 \cite{Helliwell,Linet_1986,Frolov_1987,Sakhni_1992}
 only for minimal and conformal coupling, while for other values of $\xi$ they do not coincide already in the first (in $\bs$) order.
 We refer this discrepancy
to the missing of the $\xi-$correction \textit{inside} the Green's
function. In computation of ${\langle T_{MN}\rangle}_{\ren}$ to the first order, this difference is
reflected in  terms
 $T_{MN}^{(0)}\bigl[\delta{\cal{L}}^{(\xi)}\bigr]$ and $T_{MN}^{(\xi)}\bigl[\delta{\cal{L}}^{(\xi)}\bigr]$.
 If to ignore  these two in our scheme and retain
 the two remaining in (\ref{T012}), one would obtain the old answer.


 We have shown that up to first order, in our Fourier-transform language  the results for the cosmic string spacetime
 can be obtained as the smooth limit
 of corresponding results for global monopole.
From this framework, it represents the problem of independent interest, whether this coincidence
 takes place only in the linear-in-$\bs$ order, or being the first non-vanishing part of the nonperturbative limit.

Finally, the usage of the  Perturbation Theory restricts the applicability
 by the requirement on  smallness of the angular
deficit. However, this approach is relatively simple (to the order
under consideration) and allows to take an advantage of
well-developed in QFT methods. In result, it allowed to obtain the
final expressions valid for arbitrary \mbox{$2\leqslant n\leqslant
(d-1)$} and \mbox{$d\geqslant 3$}, which, in its turn, verified
the particular cases also, what helped to justify/fix the
corresponding known results.

\bigskip

\textbf{Acknowledgment.} Yuri V.\,Grats thanks prof.
A.\,V.\,Borisov for fruitful discussions. The work of Pavel Spirin
is   supported by the RFBR grant 14-02-01092. Also PS acknowledges prof. T.\,N.\,Tomaras and grateful
to the former non-profit <<Dynasty>> foundation (Russian Federation).


\appendix

\section{Basic integrals}\label{App_ints}
Here we give the derivation of  basic integrals in the
dimensional regularization scheme we use. Such a scheme is
somewhat common in QFT but unusual in the classical theory, so it
may be instructive to briefly derive useful integrals. The
integrals are well-defined for Euclidean propagators (with
imaginary time), and  analytically generalized for the case of
Minkowski metric.
Here we rewrite the
Fourier-transforms (\ref{Four1}) in $d$ dimensions:
\begin{align}\label{Four}
{\mathcal{F}}\!\left[|\mathbf{r}|^{-\lambda}\right]\!(k)=2^{d-\lambda}\,\pi^{d/2}
\frac{\Gamma\!\left(\frac{d-\lambda}{2}\right)}{\Gamma\left(\lambda/2
\right)}\frac{1}{|\mathbf{k}|^{d-\lambda}}\,,
\end{align}
 implying the Euclidean scalar products inside.

\medskip

 The scalar single-propagator integral is defined as
\begin{equation}\label{r0}
  J^{(1)}= \int\frac{d^{\kn d} p}{(2\pi)^d}
\frac{1}{p^{\smhsp 2}-i\varepsilon}\,.
\end{equation}
Hereafter the right superscript   labels the number
of propagators.
Passing to the spherical coordinates, one obtains the integral
transformation with kernel $p^{n-3}$ acting on test function <<1>>.
Treating it as generalized function,  in \cite{Shilov} it is shown
that the latter equals zero \textit{in distributional sense}, as
well as
\begin{equation}\label{r1}
 J^{(1)}_{i_1\pp i_k} \equiv \int\frac{d^{\kn d} p}{(2\pi)^d}
\frac{p_{i_1}\pp p_{i_k}}{p^{\kn 2}-i\varepsilon}=0\,.
\end{equation}
 As it is well-known, this value is
advocated as the absence of the parameter, upon which $J^{(1)}$
could depend explicitly, since the only variable $p$ in the
integrand is integration one.

\medskip

 The
scalar two-propagator integral is defined as
\begin{equation}\label{r2}
  J^{(2)}= \int\frac{d^{\kn d} p}{(2\pi)^d}
\frac{1}{[p^{\kn 2}-i\varepsilon][\kn(p+q)^{2}-i\varepsilon]}\,.
\end{equation}
After the Wick rotation $p_0=ip_E, q_0=iq_E$ we have the analogous Euclidean integral.
Thus consider the two  following integrals  \textit{with Euclidean scalar product}:
\begin{equation}\label{ghh}
  J(q)  \equiv  \int\frac{d^{\kn d} p}{(2\pi)^d}
\frac{1}{p^{\kn 2}\kn(p+q)^{2}}\,, \qquad\qquad     I \equiv \int\frac{d^{\kn d} p \,d^{\kn d} q}{(2\pi)^{2d}}
\frac{\e^{ix(p+q)}}{p^{\kn 2} q^{2}}\,,
\end{equation}
with $ J^{(2)}=i J(q)$, indeed. Being split on the
product of identical integrals, $I$ equals
$$I=(I_0)^2\,, \qquad\qquad    I_0\equiv\int\frac{d^{\kn d} p}{(2\pi)^{d}}
\frac{\e^{ipx}}{p^{\kn2}}\,.$$ Making use of Fourier-transform
(\ref{Four}), $I$ is given by
\begin{equation}\label{ghh2}
I=\frac{1}{16\pi^d} \frac{\Gamma^2\!\left(\frac{d-2}{2}
\right)}{R^{2(d-2)}}\,, \qquad\qquad R\equiv \sqrt{x^2}.
\end{equation}
Now change variable $q\to p+q$  in $I$ (\ref{ghh}):
\begin{equation}\label{ghh3}
I=\int\frac{d^{\kn d} p \,d^{\kn d} q}{(2\pi)^{2d}} \frac{\e^{ixq}}{p^{\kn
2} (p+q)^{2}}=   {\mathcal{F}}^{-1}\!\left[ J (q) \right]\!(x)\,.
\end{equation}
Thus substituting (\ref{ghh3}) into  (\ref{ghh2}) and taking the
direct Fourier-transform with help of (\ref{Four}), $J^{(2)}(q)$
equals
\begin{equation}\label{ghh4}
 J (q)=\frac{\Gamma^2\!\left(\frac{d-2}{2}
\right)}{16\pi^d}   {\mathcal{F}}\!\left[\ir^{-2(d-2)}
\right]\!(q)=\frac{\Gamma^2\!\left(\frac{d-2}{2}
\right)\,\Gamma\!\left(-\frac{d-4}{2}\right)}{(4\pi)^{d/2}\,
\Gamma(d-2)}\,|\mathbf{q}|^{{d-4}}   .
\end{equation}
Restoring the Minkowskian $q^0$, after the $\Gamma-$function transformations the $J^{(2)}$ is given by
\begin{equation}\label{r3}
J^{(2)}=-i\frac{2\,(d-1)}{(4\pi)^{d/2}}\frac{\Gamma^2\!\left(d/2\right)\,\Gamma\!\left( -\frac{d-2}{2}\right)}{\Gamma(d)}\,
\left(q^2\right)^{\akn d/2-2 } .
\end{equation}
In the form (\ref{r3}) the arguments of all Gamma-functions
do not intersect zero at \mbox{$n\geqslant 3$}. This result is identical to the ones given in a series of QFT textbooks and derived via Feynman parametrization, but remarkably, we have used just the single basic Fourier-integral (\ref{Four}).

\noindent $\bullet$\;\,Vectorial one
reads
\begin{equation}\label{r4}
J_{M}^{(2)}= \int\frac{\ddp}{(2\pi)^d} \frac{p_{M}}{[p^{\kn 2}-i\varepsilon][\kn(p+q)^{2}-i\varepsilon]}
\end{equation}
Obviously, the result has to be proportional to $q_{M}$ as to the only
available input vector in the problem-at-hand: $J_{M}^{(2)}(q)= A_1(q)\kn q_{M}$. Contracting
this equality with $q^{M}$ and representing $p\cdot q=(1/2)
\left[(p+q)^2-p^2-q^2\right]$, one uses (\ref{r1}) and (\ref{r3})
to determine the scalar $A_1$. Thus the result turns out to be
\begin{equation}\label{r5}
J_{M}^{(2)}=-\frac{1}{2}\,
J^{(2)}q_{M}=i \frac{d-1}{(4\pi)^{d/2}}\frac{\Gamma^2\akn (d/2)\,\Gamma\!\left(- \frac{d-2}{2}\right)}{\Gamma(d)}\,
\left(q^2\right)^{\akn d/2-2 } q_{M}\,.
\end{equation}

\noindent $\bullet$\;\,Tensorial integral  is defined as
\begin{equation}\label{r6}
J_{MN}^{(2)}= \int\frac{\ddp}{(2\pi)^d} \frac{p_{M}p_{N}}{[p^{\kn 2}-i\varepsilon][\kn(p+q)^{2}-i\varepsilon]}\,.
\end{equation}
Being the symmetric 2-rank tensor, the latter should be
expressible via the flat metric $ \eta_{MN}$ and $q_{M}q_{N}-$monomial: $J_{MN}^{(2)}=A_2 \hsp q^2\kn \eta_{MN}+A_3 \hsp q_{M}q_{N}$.
 Taking the trace and substituting (\ref{r1}), one obtains the
relation
$n A_2 + A_3  =0\,.$
Projecting (\ref{r6}) on $q^{N}$ and realizing the same strategy
as before, one gets the second constraint:
$$\left(A_2 + A_3 \right)  q_{M}=\frac{1}{2}\,  J_{M}^{(2)}=\frac{1}{4 }\,  q_{M}J^{(2)}\,.$$
Resolving these two, the value of integral (\ref{r6}) is given by
\begin{align}\label{r7}
J_{MN}^{(2)}= - \frac{J^{(2)}}{4(d-1)}\,\left(q^2\eta_{MN}-d\hsp{}q_{M}q_{N}\right)
\end{align}
with trace $\eta^{MN}J_{MN}^{(2)}=0  \,.$

\noindent $\bullet$\;\,Appealing to the same computational arguments, the three-  and four-index
 integrals
\begin{equation}\label{r8}
J_{MNK}^{(2)} \equiv \int\frac{\ddp}{(2\pi)^d} \frac{p_{M}p_{N}p_{K}}{[p^{\kn 2}-i\varepsilon][\kn(p+q)^{2}-i\varepsilon]}\,,
 \qquad\qquad J_{MNKL}^{(2)}= \int\frac{\ddp}{(2\pi)^d} \frac{p_{M}p_{N}p_{K} p_{L}}{[p^{\kn 2}-i\varepsilon][\kn(p+q)^{2}-i\varepsilon]}
\end{equation}
with help of symmetry and some combinatorics, are given by
\begin{equation}\label{r9}
J_{MNK}^{(2)}= -\frac{J^{(2)}}{8\kn(d-1)}\left[(d+2) \hsp q_{M}q_{N}q_{K}- q^2  \left(q_{M} \eta_{NK}+q_{N} \eta_{MK}+q_{K} \eta_{MN}\right)\vp\right]
\end{equation}
and
 \begin{align}
 J_{MNKL}^{(2)} =\frac{J^{(2)}}{16\kn(d^{\kn 2}-1)} & \left[\vp-(d+2)\, q^2 \left(q_{M}q_{N} \eta_{KL}+ q_{M}q_{K} \eta_{NL}
 + q_{M}q_{L} \eta_{NK}+ q_{N}q_{K} \eta_{ML} + q_{N}q_{L} \eta_{MK}  +q_{K}q_{L} \eta_{MN} \vphantom{d^d_q}\right) \right.  \nn \\
 & \left.+ \vp(d+4)(d+2)\kn  q_{M}q_{N}q_{K} q_{L}+\left( q^2\right)^2 \left(\eta_{MN} \eta_{KL}+
 \eta_{MK} \eta_{NL}
 + \eta_{ML} \eta_{NK} \vphantom{d^d_q}\right)\right],
 \end{align}
respectively.

\end{document}